\documentclass[
 reprint,
superscriptaddress,
nofootinbib,
nobibnotes,
 amsmath,amssymb,
 aps,
prd,
letterpaper,
longbibliography
]{revtex4-1}

    \makeatletter
\def\@fnsymbol#1{\ensuremath{\ifcase#1\or * \or \dagger \or \ddagger
   \or \S \or \P \or ** \or \dagger\dagger
  \or \ddagger\ddagger \or \S\S \or \P\P \else\@ctrerr\fi}}
    \makeatother

\makeatletter
\def\clearfmfn{\let\@FMN@list\@empty}    
\makeatother

\usepackage{multirow}
\usepackage{graphicx}
\usepackage{dcolumn}
\usepackage{bm}
\usepackage{hyperref}
\usepackage{tabularx}
\usepackage{subfigure}

\usepackage[normalem]{ulem}
\usepackage[export]{adjustbox}
\usepackage{lipsum}
\usepackage{url}
\usepackage{mathtools}
\hypersetup{
    pdfnewwindow=true,    
    colorlinks=true,      
    linkcolor=blue,       
    citecolor=blue,       
    filecolor=blue,       
    urlcolor=blue         
}
\usepackage{orcidlink}

\usepackage{layouts}

\usepackage{scalerel}
\usepackage{accents}
\newlength{\dhatheight}

\newcommand{\blue}[1]{\textcolor{black}{#1}}

\usepackage{natbib}

\begin{document}

\title{Long-Exposure NuSTAR Constraints on Decaying Dark Matter in the Galactic Halo}

\author{Brandon~M.~Roach~\orcidlink{0000-0003-3841-6290}}
\email{roachb@mit.edu}
\affiliation{Department of Physics, Massachusetts Institute of Technology, Cambridge, Massachusetts 02139, USA}

\author{Steven~Rossland~\orcidlink{0000-0001-6304-1035}}
\email{u1019304@utah.edu}
\affiliation{Department of Physics and Astronomy, University of Utah, Salt Lake City, Utah 84112, USA}

\author{Kenny~C.~Y.~Ng~\orcidlink{0000-0001-8016-2170}}
\email{kcyng@phy.cuhk.edu.hk}
\affiliation{{Department of Physics, The Chinese University of Hong Kong, Sha Tin, Hong Kong, China}}

\author{Kerstin~Perez~\orcidlink{0000-0002-6404-4737}}
\email{kmperez@mit.edu}
\affiliation{Department of Physics, Massachusetts Institute of Technology, Cambridge, Massachusetts 02139, USA}

\author{John~F.~Beacom~\orcidlink{0000-0002-0005-2631}}
\email{beacom.7@osu.edu}
\affiliation{Center for Cosmology and AstroParticle Physics (CCAPP), Ohio State University, Columbus, Ohio 43210, USA}
\affiliation{Department of Physics, Ohio State University, Columbus, Ohio 43210, USA}
\affiliation{Department of Astronomy, Ohio State University, Columbus, Ohio 43210, USA} 

\author{Brian~W.~Grefenstette~\orcidlink{0000-0002-1984-2932}}
\email{bwgref@srl.caltech.edu}
\affiliation{Space Radiation Laboratory, California Institute of Technology, Pasadena, California 91125, USA}

\author{Shunsaku~Horiuchi~\orcidlink{0000-0001-6142-6556}}
\email{horiuchi@vt.edu}
\affiliation{Center for Neutrino Physics, Department of Physics, Virginia Tech, Blacksburg, Virginia 24061, USA}
\affiliation{Kavli IPMU (WPI), UTIAS, The University of Tokyo, Kashiwa, Chiba 277-8583, Japan}

\author{Roman~Krivonos~\orcidlink{0000-0003-2737-5673}}
\email{krivonos@iki.rssi.ru}
\affiliation{Space Research Institute (IKI), Russian Academy of Sciences, Moscow 117997, Russia}
\affiliation{Institute for Nuclear Research, Russian Academy of Sciences, Moscow 117312, Russia}

\author{Daniel~R.~Wik~\orcidlink{0000-0001-9110-2245}}
\email{wik@astro.utah.edu}
\affiliation{Department of Physics and Astronomy, University of Utah, Salt Lake City, Utah 84112, USA}

\date{\today}

\begin{abstract}
We present two complementary NuSTAR x-ray searches for keV-scale dark matter decaying to monoenergetic photons in the Milky Way halo. In the first, we utilize the known intensity pattern of unfocused stray light across the detector planes---the dominant source of photons from diffuse sources---to separate astrophysical emission from internal instrument backgrounds using ${\sim}$7- Ms/detector deep blank-sky exposures. In the second, we present an updated parametric model of the full NuSTAR instrument background, allowing us to leverage the statistical power of an independent ${\sim}$20-Ms/detector stacked exposures spread across the sky. Finding no evidence of anomalous x-ray lines using either method, we set limits on the active-sterile mixing angle $\sin^2(2\theta)$ for sterile-neutrino masses 6--40 keV. The first key result is that we strongly disfavor a ${\sim}$7-keV sterile neutrino decaying into a 3.5-keV photon. The second is that we derive leading limits on sterile neutrinos with masses ${\sim}$15--18 keV and ${\sim}$25--40 keV, reaching or extending below the big bang nucleosynthesis limit. In combination with previous results, the parameter space for the neutrino minimal standard model is now nearly closed.

\end{abstract}

\maketitle

\section{\label{sec:intro} Introduction}

\par Nearly a century of cosmological observations have indicated the presence of gravitating degrees of freedom that do not couple to electromagnetism with the same strength as the visible Standard Model (SM) particles. One class of searches for this dark matter (DM, hereafter symbolized $\chi$) is indirect detection, in which astrophysical observatories are used to search for the decay and/or annihilation of DM particles into stable SM particles (see, e.g., Refs.~\cite{Funk:2013gxa,Gaskins:16,PerezdelosHeros:2020qyt}). Unlike charged-particle cosmic rays, photons and neutrinos are not scattered by astrophysical magnetic fields, allowing any putative DM signal to be correlated against known astrophysical sources. 
\par A popular DM candidate with a final-state photon signal amenable to indirect detection is the keV-scale sterile neutrino. Models such as the neutrino minimal standard model ($\nu$MSM, Refs.~\cite{Asaka:2005an,Asaka:2006nq,Shaposhnikov:2006xi,Laine:2008pg}) incorporate these sterile neutrinos while simultaneously seeking to account for the observed neutrino mass spectrum and the cosmological matter/antimatter asymmetry. Such sterile neutrinos are particularly interesting as a candidate for indirect DM searches, as their radiative decays $\chi \rightarrow \gamma\nu_\ell$ to the SM lepton neutrinos $\nu_\ell$ would produce a monoenergetic x-ray line with energy $E_\gamma = m_\chi/2$ and decay rate $\Gamma_{\chi\rightarrow \gamma \nu}$ set by $m_\chi$ and the active-sterile mixing $\sin^2(2\theta)$. In the early Universe, sterile neutrinos may have been produced via oscillation-induced mixing with the SM neutrinos \cite{Dodelson:1993je}, with a primordial lepton asymmetry potentially enhancing the rates \cite{Shi:1998km}. 
\par Many space-based x-ray observatories have contributed to the search for radiative sterile-neutrino DM decays~\cite{Boyarsky:2005us,Boyarsky:2006fg,Boyarsky:2007ge,Watson:2006qb,Yuksel:2007xh,Loewenstein:2008yi,RiemerSorensen:2009jp,Horiuchi:2013noa,Urban:2014yda,Tamura:2014mta,Figueroa-Feliciano:2015gwa,Iakubovskyi:2015dna,Ng:2015gfa,Aharonian:2016gzq,Sekiya:2015jsa,Dessert:2018qih,Hofmann:2019ihc,Sicilian:2020a,Bhargava:2020a,Foster:2021ngm,Silich:2021sra}. These include focusing telescopes such as the Chandra X-ray Observatory (CXO), XMM-Newton, Hitomi, and Suzaku, and nonfocusing instruments such as Halosat, the Fermi Gamma-ray Burst Monitor and the Spectrometer aboard INTEGRAL. The different sensitivity bands of these instruments led to a gap in the mass-mixing angle parameter space for sterile-neutrino masses ${\sim}$10--25 keV. Additionally, claims of the detection of an anomalous x-ray line at ${E_\gamma \simeq 3.5 \text{\,keV}}$ ($m_\chi \simeq 7\text{\,keV}$, Refs.~\cite{Bulbul:2014sua,Boyarsky:2014jta}) motivate covering the sterile-neutrino DM parameter space with as many instruments, observation targets, and analysis techniques as possible.
\par The NuSTAR observatory (launched in 2012, Ref.~\cite{Harrison:2013}) is uniquely suited to fill in this gap in the mass-mixing-angle parameter space, and to test the origin of the 3.5-keV anomaly. Following a search for sterile-neutrino DM in focused observations of the Bullet cluster \cite{Riemer-Sorensen:2015kqa}, subsequent NuSTAR analyses used the so-called 0-bounce (unfocused stray light) photons to derive leading limits on sterile-neutrino decays in blank-sky extragalactic fields \cite{Neronov:2016wdd}, the Galactic Center, \cite{Perez:2016tcq}, the M31 galaxy \cite{Ng:2019a}, and the Galactic bulge \cite{Roach:2019ctw}, though these analyses were limited by a combination of astrophysical background emission, limited statistics, and systematic deviations from the fiducial instrument background model. 
\par In this paper, we leverage NuSTAR's extensive observational catalog since 2012 to derive robust constraints on $\sin^2(2\theta)$ across much of the 6--40 keV mass range, using two independent datasets and analysis techniques. First, we use the known intensity pattern of unfocused stray light on the NuSTAR detectors to separate astrophysical emission from internal instrument backgrounds using 7-Ms/detector deep blank-sky exposures, which allows us to derive a novel limit on $\sin^2(2\theta)$ for sterile-neutrino masses between 6--40 keV. This technique allows us to greatly suppress instrument backgrounds, and especially to probe the challenging ${\sim}$6--10-keV mass range. Second, we apply an improved parametric model of the NuSTAR instrumental and astrophysical backgrounds to the full ${\sim}$20-Ms/detector dataset, providing improved sensitivity at higher masses, ${\sim}$25--40\,keV.
\par In Sec.~\ref{sec:nustar}, we briefly describe the aspects of the NuSTAR observatory design relevant to our sterile-neutrino search. In Sec.~\ref{sec:spatial}, we describe the novel ``spatial-gradient'' technique that allows us to separate 0-bounce photons from detector backgrounds. In Sec.~\ref{sec:parametric}, we discuss the development and implementation of the updated NuSTAR parametric background model. In Sec.~\ref{sec:constraints}, we scan the NuSTAR spectra from both analysis techniques for evidence of decaying DM, particularly keV-scale sterile neutrinos. We conclude in Sec.~\ref{sec:conclusions}.

\section{\label{sec:nustar}NuSTAR As A Dark-Matter Observatory}
The aspects of the NuSTAR instrument relevant for sterile-neutrino searches have been described in previous analyses \cite{Neronov:2016wdd,Perez:2016tcq,Ng:2019a,Roach:2019ctw}; here, we reiterate the most important points. 
\subsection{\label{sec:nustar_optics}NuSTAR Optics Modules}
NuSTAR carries two coaligned x-ray telescopes labeled A and B, each comprised of an optics module (OM) and a focal plane module (FPM) separated by the observatory's 10-meter carbon-fiber mast. The OMs are conical approximations to the Wolter-I grazing-incidence design, with properly focused x-rays reflecting twice inside the OMs---first against the parabolic mirrors and second against the hyperbolic mirrors---hence their alternative name of ``2-bounce'' (2b) photons. The multilayer construction of the mirrors with alternating layers of platinum/silicon carbide and tungsten/silicon affords NuSTAR considerable focused area for photon energies $E_\gamma$ between 3--79 keV. The focused FOV of NuSTAR subtends a solid angle ${\Delta \Omega_\text{2b} = 13^\prime \times 13^\prime \approx 0.047\text{\,deg}^2}$, and the optics provide a maximum FOV-averaged effective area ${\langle A_\text{2b} \rangle \approx 170 \text{\,cm}^2}$ per FPM for photon energies ${E_\gamma \approx 10 \text{\,keV}}$. Thus, the maximum 2-bounce grasp $\langle A_\text{2b} \Delta \Omega_\text{2b}\rangle \approx 8 \text{\,cm}^2\,\text{\,deg}^2$ per FPM.
\subsection{\label{sec:nustar_fpms}NuSTAR Focal Plane Modules}
\par At the opposite end of the mast from the optics modules sit the  FPMs. Each FPM consists of a solid-state detector array, a cesium iodide anticoincidence shield to veto incoming cosmic rays, a series of three annular aperture stops to block off-axis photons from striking the detectors, and a ${\sim}0.1$-mm beryllium window with energy-dependent transmission coefficient $\mathcal{E}_\text{Be}$ to block lower-energy photons. Each FPM detector array consists of four cadmium zinc telluride (CdZnTe) crystals, with each crystal having dimensions ${(20\times20\times2)\,\text{mm}^3}$ and segmented into a $32\times 32$ grid of $(0.6\text{\,mm})^2$ pixels. The detectors have energy resolution ${\sim}0.4\text{\,keV}$ FWHM for photon energies $E_\gamma \lesssim 10\text{\,keV}$, increasing to ${\sim}0.9\text{\,keV}$ FWHM at $E_\gamma \approx 80 \text{\,keV}$. The detector response is defined for energies 1.6--164 keV and is divided into 4096 channels of width 40 eV. At present, only the $E_\gamma > 3 \text{\,keV}$ response is known with sufficient precision for this study, but work is ongoing to extend to lower energies \cite{Grefenstette:2018a}.
\par Before interacting with the active CdZnTe, photons must pass through the ${\sim}$100-nm platinum contact, as well as a ${\sim}$200-nm ``dead layer'' of inactive CdZnTe. The thickness of these layers vary somewhat between individual detector chips, calibrated using extensive off-axis observations of the Crab nebula \cite{Madsen:2015,Madsen:2021hjz}. We incorporate the energy-dependent throughput of the platinum contact and CdZnTe dead layer into an overall transmission coefficient $\mathcal{E}_\text{det}$.
\subsection{\label{sec:nustar_zerobounce}The 0-Bounce Technique}
\par Unlike previous focusing x-ray observatories such as CXO or XMM-Newton, the path between the NuSTAR optics bench and the detector plane is largely open to the sky. This configuration allows photons with off-axis angles ${\sim}$1--3$^\circ$ to strike the detectors without being focused by the mirror optics, hence their name ``0-bounce'' (0b) photons. The effective 0-bounce solid angle $\Delta \Omega_\text{0b} \equiv \int_\text{FOV} \xi_\text{0b}\,d\Omega$ is determined by the geometry of the aperture stops within the FPMs, partially blocked by the optics bench to form a crescent ``Pac-Man'' gradient in efficiency $\xi_\text{0b}$ across the detectors (see, e.g., Refs.~\cite{Wik:2014,Perez:2016tcq,Krivonos:2020qvl} for a schematic). Since these 0-bounce photons bypass the focusing optics, the unfocused effective area $A_\text{0b}$ is limited by the physical ${\sim}13\text{\,cm}^2$ area of each detector array. The usable area of each detector chip ranges between 3.12--3.19 $\text{cm}^2$ due to the varying amount of ``bad pixels'' flagged in the Calibration Database (CALDB). As detector arrays A/B have different orientations with respect to the optics bench, the 0-bounce efficiency and effective solid angle $\Delta \Omega_\text{0b}$ vary across the detector chips, and between FPMA and FPMB. Taking ${\Delta \Omega_\text{0b} \approx 4.5\text{\,deg}^2}$ as the approximate solid angle for each FPM, the 0-bounce grasp ${\langle A_\text{0b} \Delta \Omega_\text{0b} \rangle \approx 55 \text{\,cm}^2\text{\,deg}^2}$ per FPM, nearly an order of magnitude larger than the maximum 2-bounce grasp. Additionally, unlike the 2-bounce grasp, the 0-bounce grasp is essentially constant for $E_\gamma \gtrsim 10$\,keV. These 0-bounce photons are ideal for studies of diffuse x-rays on ${\sim}$degree scales, e.g., dark matter decay.
\subsection{\label{sec:nustar_expected}Expected DM Signal}
With the NuSTAR instrument responses in hand, we may readily calculate the expected DM-decay-induced photon intensity $\mathcal{I}_\text{DM} \equiv d^2 F_\gamma/dE_\gamma d\Omega$ at the telescope:
\begin{align}
    \mathcal{I}_\text{DM} &=  \frac{\Gamma}{4\pi m_\chi} \frac{dN}{dE_\gamma} \left[\frac{1}{\Delta \Omega} \int_\text{FOV}\xi \,d\Omega \int_\text{LOS} \rho_\chi \,ds \right] \nonumber \\
    &= \frac{\Gamma}{4\pi m_\chi} \frac{dN}{dE_\gamma} \left[ \frac{1}{\Delta \Omega} \int_\text{FOV} \xi \frac{d\mathcal{D}}{d\Omega}\,d\Omega \right] \\
    &= \frac{\Gamma}{4\pi m_\chi} \frac{dN}{dE_\gamma}\left\langle \frac{d\mathcal{D}}{d\Omega}\right\rangle .\nonumber
\end{align}
Here, $\Gamma$ is the decay rate to some final state with associated photon spectrum $dN/dE_\gamma$, the latter being normalized to the number of final-state photons in that channel. For the two-body final states with $m_x \ll m_\chi$ considered in this work (i.e., $\chi \rightarrow \gamma x$) and assuming the linewidth is much less than the ${\sim}0.4$-keV FWHM detector energy resolution, $dN/dE_\gamma \simeq \delta(E_\gamma - m_\chi/2)$. The term in square brackets is the FOV-averaged DM column density per solid angle $\langle d\mathcal{D}/d\Omega\rangle$, where $\Delta \Omega = \int_\text{FOV} \xi\,d\Omega$ is the effective solid angle for the 0-bounce or 2-bounce FOV, as appropriate, and $s$ is the distance along the line of sight (LOS) through the halo. To convert to the measured count rate $d^2 N/dE_\gamma dt$, we evaluate $\mathcal{I}_\text{DM}$ for the 0-bounce and 2-bounce apertures, and fold $\mathcal{I}_\text{DM}$ with the appropriate solid angles, effective areas, and detector response matrices.
\par
\begin{figure}[t!]
    \centering
    \hspace*{-1cm}
    \includegraphics[scale=0.9]{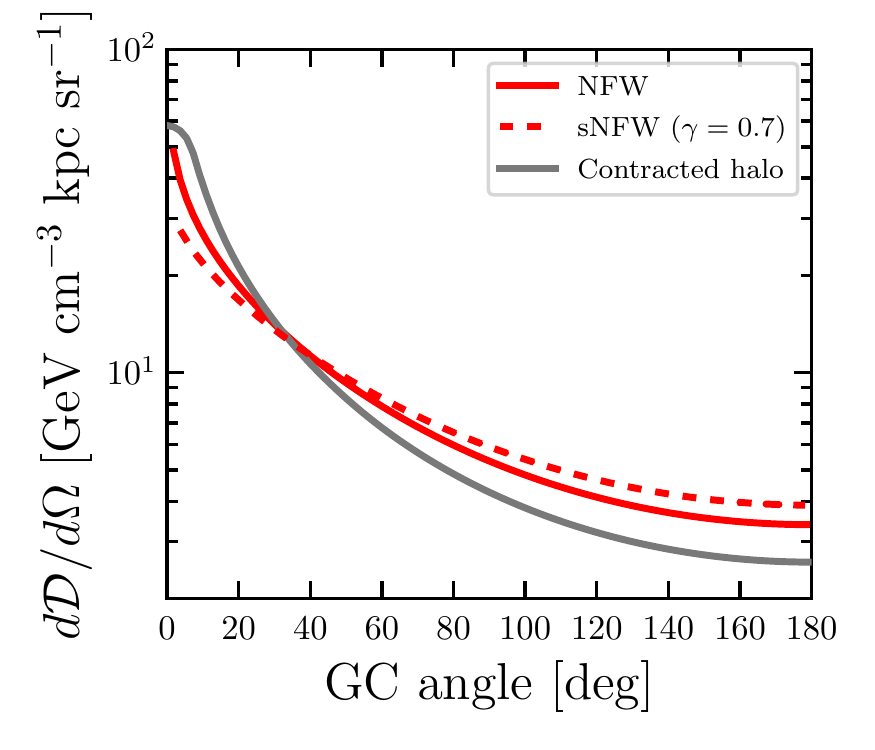}
    \caption{DM column density $d\mathcal{D}/d\Omega \equiv \int_\text{LOS} \rho_\chi\,ds$ versus angle from the Galactic Center (GC) for NFW and shallow NFW (sNFW) profiles, as well as the contracted halo of Ref.~\cite{Cautun:2019eaf}. Most NuSTAR observations in this work are at moderate distances from the GC, minimizing profile uncertainties. See Sec.~\ref{sec:nustar_expected} for further details.}
    \label{fig:dfactors}
\end{figure}

 The choice of DM density profile $\rho_\chi$ as a function of galactocentric distance $r$ is an important consideration for indirect DM searches. One popular choice of profile is the generalized Navarro-Frenk-White profile ${\rho_\text{gNFW} \propto (r/r_s)^{-\gamma}[1+(r/r_s)]^{\gamma -3}}$, where $r_s$ is the scale radius \cite{Zhao:1995cp}. We consider a canonical DM-only NFW profile with $\gamma = 1$ \cite{Navarro:1996gj} as well as a shallow (sNFW) profile with $\gamma=0.7$ \cite{Pato:2015dua,Karukes:2019jxv}, with both NFW variants having local DM density ${\rho_\chi(r_\odot) = 0.4\text{\,GeV\,cm}^{-3}}$ \cite{deSalas:2019pee,deSalas:2020hbh,Sofue:2020rnl}. Finally, we consider the contracted Milky Way halo model of Ref.~\cite{Cautun:2019eaf} with local DM density ${\rho_\chi(r_\odot) \approx 0.3 \text{\,GeV\,cm}^{-3}}$, though since this model is only validated for $r > 1\text{\,kpc}$, we conservatively assume that the DM density within $r<1\text{\,kpc}$ is constant. For all Galactic DM profiles, we adopt ${r_\odot = 8.1 \text{\,kpc}}$ for the Sun's galactocentric distance \cite{Bobylev:2021zzn}. The DM column density as a function of viewing angle from the Galactic Center (GC) is shown in Fig.~\ref{fig:dfactors}. So that we may set conservative upper limits on the DM decay rate, we do not include enhancements to the DM column density either from extragalactic sources or from possible substructure in the Milky Way (MW) halo. The impact of different profile choices on our DM decay limits is discussed in Secs.~\ref{sec:constraints_spatial} and ~\ref{sec:constraints_parametric}.

\section{\label{sec:spatial}NuSTAR Spatial Gradient Analysis}
In this section, we describe a novel application of the NuSTAR 0-bounce technique to dark-matter searches to stacked ${\sim}$7-Ms/detector exposures of blank-sky fields: using the known spatial gradient of 0-bounce photons on the detectors to separate instrumental backgrounds from astrophysical x-ray emission.

\subsection{\label{sec:roman_observations}NuSTAR Observations and Data Processing}
The NuSTAR dataset used in our spatial-gradient analysis was previously analyzed in a study of the cosmic x-ray background (CXB, Ref.~\cite{Krivonos:2020qvl}); here, we review several key aspects. The observations were conducted from 2012--2016 as part of the NuSTAR extragalactic survey program of the COSMOS~\cite{Civano:2015hnp}, EGS~\cite{Davis:2006tn}, ECDFS~\cite{Mullaney:2015qgo}, and UDS~\cite{Masini:2018xsj} blank-sky fields. Initial data reduction was performed with \textsc{nustardas} \textsc{v}1.8.0, with the flags \texttt{SAAMODE=strict} and \texttt{TENTACLE=yes} used to exclude NuSTAR passages through the South Atlantic Anomaly (SAA). A threshold of 0.17 counts s$^{-1}$ in the 3--10 keV range on FPMA and FPMB was used as a threshold for excluding observations due to heightened solar and/or geomagnetic activity. Following these cuts, the total cleaned exposure time for the NuSTAR detectors is ${\sim}7$ Ms/FPM, shown in Sec.~\ref{sec:parametric}. We do not exclude any detector regions corresponding to known astrophysical x-ray
sources, as these sources tend to be few in number and faint in comparison to the unresolved CXB; instead, we allow any faint sources in the FOV to contribute to the 0-bounce spectrum. 
\par
\begin{figure}[b]
    \centering
    \hspace*{-1.5cm}

    \includegraphics[scale=0.9]{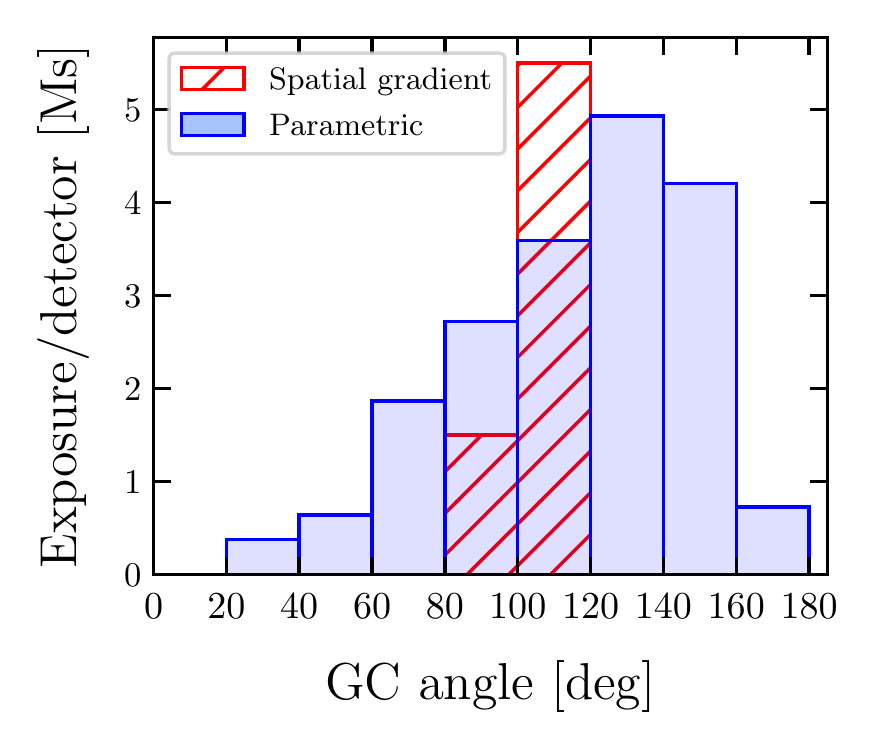}
    \caption{Average cleaned exposure time per detector for the spatial-gradient analysis of Sec.~\ref{sec:spatial} (red hatched) and the parametric analysis of Sec.~\ref{sec:parametric} (blue shaded) versus angular distance from the Galactic Center.} 
    \label{fig:exposures}
\end{figure}

 The average exposure time per detector versus angular distance from the GC is shown in Fig.~\ref{fig:exposures} for both analyses described in this work. The extragalactic survey fields included in the present spatial-gradient analysis are located at similar distances from the GC (${\sim}$95--110$^{\circ}$), and thus share similar DM column densities. Finally, the high latitudes of these fields ($|b|\sim\,$40--60$^\circ$) place them far from x-ray line or continuum emission in the Galactic plane.

\subsection{\label{sec:roman_data}Spatial-Gradient Analysis}
The crux of the spatial-gradient analysis is the fact that photons observed by NuSTAR have different spatial geometries when considered in detector coordinates. First, internal detector backgrounds (both line and continuum) are observed to have an essentially uniform distribution across each detector, though the overall rates between detectors may differ as a result of their different thicknesses (further discussed in Sec.~\ref{sec:parametric_model}). \blue{Second, the focused CXB component includes both 2-bounce photons from the $13^\prime \times 13^\prime$ focused FOV (whose detector gradient follows the vignetting of the optics) and 1-bounce ``ghost-ray'' photons up to $30^\prime$ off axis~\cite{10.1117/1.JATIS.3.4.044003}.} Finally, the 0-bounce photons hitting the detectors from ${\sim}\text{1--3}^\circ$ off-axis (principally from the CXB) manifest as a ``Pac-Man'' shaped gradient on the detectors. The solid angle of sky observed by each pixel (and hence the intensity pattern on the detector, assuming a uniform flux across the 0-bounce FOV) can be readily calculated from the known positions of the NuSTAR detectors, aperture stops, and optics bench using the \textsc{nuskybgd} code \cite{Wik:2014}.
\par From this, we constructed the same likelihood model as that used in Ref.~\cite{Krivonos:2020qvl}, with two spectral components: a spatially uniform internal detector component, and a 0-bounce component following the ``Pac-Man'' spatial gradient. We divide the 3--20 keV energy range into 100 bins equally spaced in $\log_{10} E_\gamma$\blue{, and bin the data using the $64\times 64$ RAW detector pixels ($32\times 32$ pixels per detector chip) to provide sufficient counts in each energy bin.} The expected total counts $N_\text{pix}$ accumulated in the $i^\text{th}$ pixel during exposure time $T$ is given by
\begin{equation}
    N_{\text{pix},i}(E_\gamma) = (C_\text{int} M_\text{int} + C_\text{0b} R_\text{pix} \mathcal{E}_\text{tot} A \Omega)_i T
\end{equation}
where $C_{\text{int},i} \equiv (dN/dt)_i$ is the internal background event rate, $M_\text{int}$ encodes the nonuniformity and differences in relative normalization between the eight detectors obtained using 10--20-keV occulted data, $C_{\text{0b},i} \equiv (dF/d\Omega)_i$ is the 0-bounce flux per solid angle, $\mathcal{E}_\text{tot}\equiv \mathcal{E}_\text{det} \mathcal{E}_\text{Be}$ is the energy-dependent transmission coefficient of the inactive detector surface layer and beryllium entrance window as described in \textsc{caldb} \textsc{v}20200813, $R_\text{pix}$ is the matrix encoding the nonuniform pixel response in the NuSTAR \textsc{caldb}, $A_i$ is the physical $(0.6\text{\,mm})^2$ area of each pixel, $\Omega_i$ is the effective 0-bounce solid angle calculated using \textsc{nuskybgd}, and $T$ is the exposure time of the observation. \blue{We do not include a focused CXB component for several reasons. First, the focused CXB signal is expected to be faint---nearly an order of magnitude fainter than the 0-bounce CXB signal (see Fig.~9 of Ref.~\cite{Wik:2014}). The focused CXB signal is thus at or below the level of the internal detector background, making it extremely challenging to detect the spatial variations in the focused CXB signal. Second, the spatial variations in the focused CXB signal are further flattened when the data are binned in RAW detector pixels. Thus, our analysis does not distinguish the focused CXB signal from the spatially flat internal detector background, so we allow the former component to be absorbed by the latter.}
\par For each energy bin, we construct the likelihood (suppressing the $E_\gamma$ dependence for clarity)
\begin{equation}
     \mathcal{L} = \prod_i \left[ \frac{N_{\text{pix}}^{\mathcal{N}} \exp[-N_{\text{pix}}] }{\mathcal{N}!}\right]_i
\end{equation}
and minimize $-2\ln\mathcal{L}$ with respect to $C_\text{int}$ and $C_\text{0b}$, where $\mathcal{N}_i$ is the observed number of counts in the $i^\text{th}$ pixel. The product runs over all pixels and NuSTAR observations. This produces nearly pure 0-bounce spectra and their corresponding detector response files for both FPMs. Modulo the narrower energy bins in this work, the spectra of Ref.~\cite{Krivonos:2020qvl} are identical to those shown here. 
\par (We note that this data processing was completed before the release of the updated \textsc{caldb} \textsc{v}20211020, which modified the 2-bounce vignetting profile, detector response matrices, and inactive CdZnTe throughput $\mathcal{E}_\text{det}$. Of these, changes in $\mathcal{E}_\text{det}$ have the greatest effect on our DM constraints; however, the variations in $\mathcal{E}_\text{det}$ between \textsc{caldb} versions are ${\sim}$5\% at $E_\gamma = 3\text{\,keV}$, with the agreement improving with increasing $E_\gamma$. In any case, this effect is subdominant compared to the ${\sim}$15--25\% DM profile uncertainties discussed in Sec.~\ref{sec:constraints_spatial}.)

\begin{figure*}
    \centering
    \includegraphics[scale=0.8]{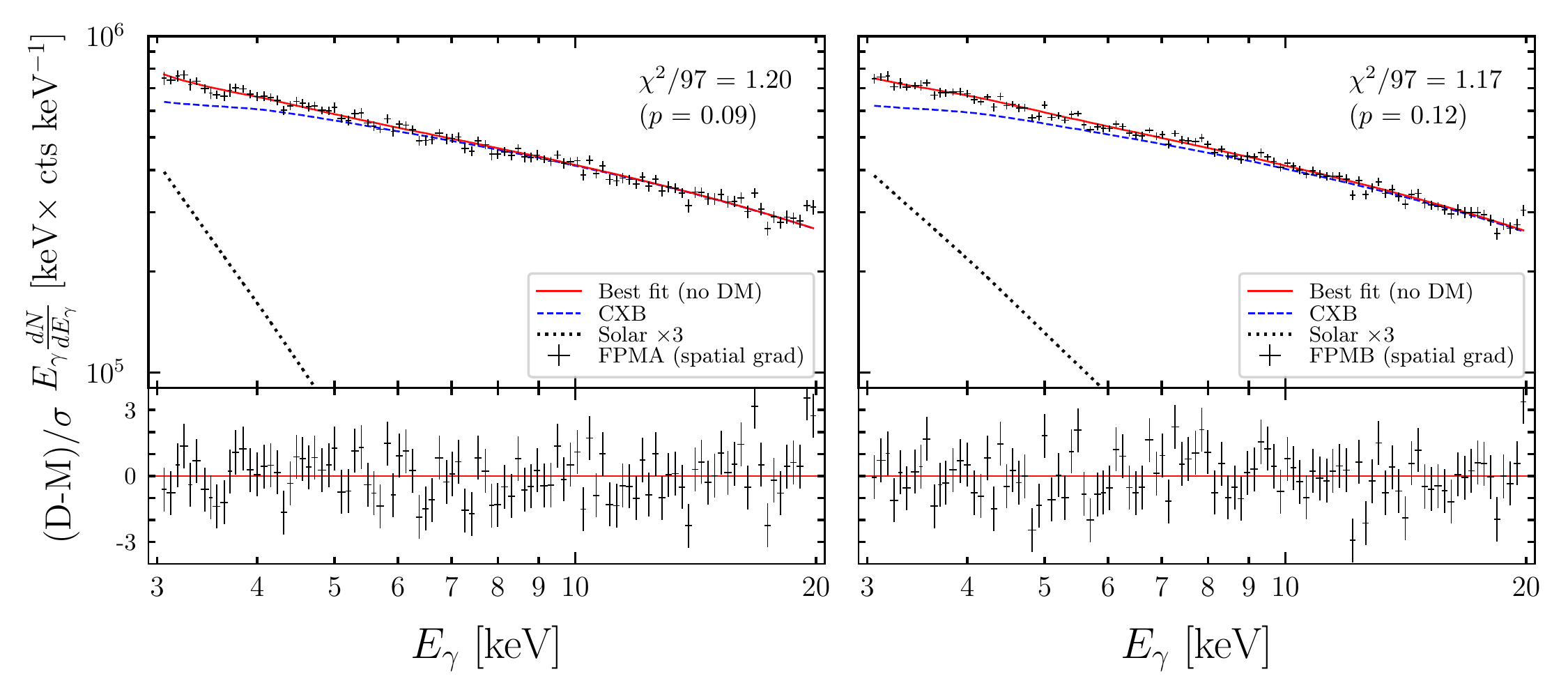}
    \caption{0-bounce spectra and best-fit models without DM, derived from the NuSTAR spatial-gradient technique for FPMA (left) and FPMB (right). The spectra $d^2 F_\gamma/dE_\gamma d\Omega$ are scaled by the nominal exposure time (7 Ms), effective area ($12.5\text{\,cm}^2$) and solid angle ($4.5\text{\,deg}^2$) for presentation. The solar power law normalization is multiplied by a factor of three for visibility. The bottom panels show the residuals (Data--Model) scaled by statistical uncertainty $\sigma$.}
    \label{fig:spatial_spectra}
\end{figure*}

\begin{table}[t!]
    \centering
    \begin{tabularx}{\columnwidth}{XXXX}
    \hline
    \hline
        &&&\\[-1em]
        Module & $\mathcal{F}^{\text{CXB}}_{3-20\text{\,keV}}$ & $\mathcal{F}_{3-20\text{\,keV}}^\text{solar}$ & $\Gamma_\text{solar}$\\
        \hline
        &&&\\[-1em]
        FPMA & $2.82^{+0.02}_{-0.02}$ & 0.07$^{+0.01}_{-0.01}$& 4.5$^{+0.5}_{-0.4}$\\
        &&&\\[-1em]
        FPMB & $2.77^{+0.03}_{-0.04}$ & $0.13^{+0.03}_{-0.02}$ & 3.3$^{+0.4}_{-0.4}$ \\
        \hline
        \hline
    \end{tabularx}
    \caption{Best-fit parameters and 68\% confidence intervals for the CXB and solar power law  for the {NuSTAR} spectra derived from the spatial-gradient analysis. The energy fluxes $\mathcal{F}_{3-20\text{\,keV}}^\text{CXB}$ and $\mathcal{F}_{3-20\text{\,keV}}^\text{solar}$ have units $10^{-11}$ erg cm$^{-2}$ s$^{-1}$ deg$^{-2}$.}
    \label{tab:krivonos_pars}
\end{table}

\subsection{\label{sec:roman_model}Spectral Model}
To fit the resulting unfolded spectra for FPMA and FPMB, shown in Fig.~\ref{fig:spatial_spectra}, we construct the model from Ref.~\cite{Krivonos:2020qvl} in \textsc{xspec} 12.11.1. The CXB intensity ${\mathcal{I}_\text{CXB} \equiv d^2 F_\gamma/dE_\gamma d\Omega}$ is parametrized by the model proposed by Ref.~\cite{Gruber:1999} for the energy range 3--60 keV, rescaled from units of sr$^{-1}$ to deg$^{-2}$ for convenience:
\begin{align}
 \mathcal{I}_\text{CXB} &= 0.0024\left(\frac{E_\gamma}{1\text{\,keV}}\right)^{-\Gamma_\text{CXB}} \exp\left[-\frac{E_\gamma}{E_\text{fold}}\right]\\
    &\hspace{80pt}\text{cm}^{-2}\,\text{s}^{-1}\,\text{\,deg}^{-2}\,\text{keV}^{-1} \nonumber
\end{align}
We adopt the canonical values $\Gamma_\text{CXB} = 1.29$ and $E_\text{fold} = 41.13\text{\,keV}$ proposed in Ref.~\cite{Gruber:1999} and shown to provide good fit quality in Ref.~\cite{Krivonos:2020qvl}. We also include an additional power law model of the form $E_\gamma^{-\Gamma_\text{solar}}$ to account for any residual solar emission particularly during the active years ${\sim}$2013--2014, with both the overall flux level and spectral index of the solar component allowed to vary (though we impose a limit $\Gamma_\text{solar}>2$ to prevent the solar component from becoming degenerate with the CXB). We do not include a model component to account for x-ray attenuation in the interstellar medium (ISM), as the equivalent neutral hydrogen column density $N_\text{H}$ in the direction of these high-latitude survey fields is small, ${\lesssim 2\times 10^{20}\text{\,cm}^{-2}}$ \cite{Dickey:1990mf,Kalberla:2005ts}. (Adopting the attenuation cross sections from Ref.~\cite{Wilms:2000ez} and solar elemental abundances from Ref.~\cite{Anders:1989zg}, this corresponds to an equivalent optical depth $\tau \sim 10^{-3}$ at 3 keV, indicating negligible ISM attenuation that further decreases with energy.)
\par With only three free parameters, we obtain good fits with $\chi^2/97 = 1.20$ for FPMA and 1.17 for FPMB ($p$-values 0.09 and 0.12, respectively). The best-fit energy fluxes $\mathcal{F}_\text{3--20\,keV} \equiv \int_3^{20}E_\gamma \mathcal{I}\,dE_\gamma$ and solar power law indices are given in Table~\ref{tab:krivonos_pars}. The CXB fluxes for FPMA and FPMB agree with each other and with the results of Ref.~\cite{Krivonos:2020qvl} at the few percent level, consistent with the expected cross calibration uncertainty between the FPMs. To ensure that our eventual DM limits are consistent with the expected statistical fluctuations, we simulate $10^3$ spectra each for FPMA and FPMB using the \textsc{xspec} tool \texttt{fakeit} by convolving the best-fit models in Table~\ref{tab:krivonos_pars} with the appropriate instrument response files and injecting Poisson noise. These mock spectra are then passed through the same analysis chain as the original spectral data in Sec.~\ref{sec:constraints}.

\section{\label{sec:parametric}NuSTAR Parametric Background Analysis}
In this section, we describe the development of an improved parametrization of the NuSTAR instrument background, and its application to ${\sim}$20-Ms/detector stacked exposures spread across the sky.

\subsection{\label{sec:allsky_data}NuSTAR Data Processing}
We considered all observations from 2012--2017, minus those from Sec.~\ref{sec:spatial} to produce a dataset independent from the spatial-gradient analysis, leaving ${\sim}2000$ observations with exposure $>{1}\text{\,ks}$. Our data processing strategy was optimized to provide as ``clean'' a spectrum as possible (i.e., minimizing contamination from astrophysical sources and geomagnetic/solar activity) while accumulating as much observation time as possible. To prevent contamination from diffuse emission and point sources in the Galactic plane, we conservatively exclude all observations with Galactic latitudes $|b|<15^\circ$. This leaves ${\sim}600$ observations to analyze, listed in \cite{github_roach}.
\par We begin the point-source removal process by creating a single 3--30 keV image per observation, smoothed with a Gaussian kernel of radius 6 pixels. Any candidate source with peak intensity greater than five times the expected background rate from \textsc{nuskybgd} is flagged and fit with a circular exclusion region. The radius of this region is determined by the peak intensity value in relation to a model point-spread function (PSF) as described in previous work \cite{Madsen:2015}. To ensure the wings of the PSF have minimal influence on the resulting spectra, we define the outer boundary of the source exclusion regions such that the source event rate falls below 3\% of the expected background rate. This creates exclusion regions many times the apparent size of the source, but due to NuSTAR's extended PSF, allows us to confidently utilize images with known sources.

\par At this stage, background light curves (excluding detected sources but still containing photons from faint 0-bounce and/or 2-bounce sources) would ideally have no temporal variation. However, SAA passages each orbit temporarily increase the detector background, particularly at energies $E_\gamma \gtrsim 50\text{\,keV}$, and enhanced solar activity can increase the background at energies $E_\gamma \lesssim 10\text{\,keV}$.
These variations occur on generally short (few-minute) timescales compared to  the ${\sim}$day-length timescales of individual NuSTAR observations.
As such, these ``flaring" periods can be readily identified as deviations from the mean background rate and removed; however, some light curves even without flaring exhibit an overall sinusoidal variation with a period ${\sim}$1 day, resulting from precession of the observatory's orbital motion with respect to the geomagnetic rigidity cutoff~\cite{bwgref}. This sinusoidal variation must be accounted for to ensure proper identification and removal of flaring events.
\par To minimize bias in the initial processing, we implement a data-driven procedure to exclude flares. Following astrophysical x-ray source exclusion, we filter the event files to include only the energy range 50--100 keV. This energy band contains many fluorescence and activation features of the NuSTAR instrument, which are particularly sensitive to flaring. 
We exclude all time intervals whose event rate is ${>3.5\sigma}$ above the expected background rate, with a second filtering step performed to exclude any low-level flares missed due to the presence of a larger flare. Finally, a source exclusion region, if it exists, is then applied to the 3--7~keV energy band where the Sun is the dominant contribution to the event rate, but still below the event rate expected from astrophysical x-ray sources. The average exposure time per detector as a function of angle from the Galactic Center is shown in Fig.~\ref{fig:exposures}.

\par At this stage, the event files have been filtered both spatially (removing astrophysical x-ray sources and creating separate event files for each detector) and temporally (removing flaring periods). From these filtered event files, we extract spectra and 2-bounce effective area curves $A_\text{2b}(E_\gamma)$ from each detector individually using \textsc{nustardas} \textsc{v}2.0.0 and calibration database (\textsc{caldb}) \textsc{v}20200813 in extended-source mode. The $A_\text{2b}$ curves incorporate the beryllium shield throughput $\mathcal{E}_\text{Be}$ but not the detector dead layer throughput $\mathcal{E}_\text{det}$, as the latter is applied later. We use \textsc{nuskybgd} to calculate the 0-bounce effective area $A_\text{0b}$ and solid angle $\Delta \Omega_\text{0b}$ for each cleaned single-detector event file. We create one stacked spectrum per detector with exposure times shown in Table~\ref{table:parametric_exposure}. The effective areas and solid angles for each stacked spectrum are the exposure-time-weighted averages of the individual observations. Note that $A_\text{0b}$ strictly decreases after masking astrophysical source regions, with the greatest reductions occurring on detectors A0 and B0. This is a result of the optical axis landing on detectors A0 and B0, so images of targeted point sources---and hence their exclusion regions---land mainly on those detectors as well (and to a lesser extent on A3 and B3; see Fig.~5 of Ref.~\cite{Harrison:2013}). In contrast, $\Delta \Omega_\text{0b}$ may either increase or decrease following region masking, depending on blocking of the pixels by the optics bench (see Fig.~2 of Ref.~\cite{Krivonos:2020qvl}).

\subsection{\label{sec:parametric_model}NuSTAR Background Model}
The NuSTAR instrument background consists of four components that vary with energy and position on the detector plane. Here, we describe the procedure used to derive a phenomenological model of the instrument background applicable to the ${\sim}20$-Ms/detector stacked spectra of Sec.~\ref{sec:allsky_data}, as well as verifying the stability of the model from 2012--2017. We note at the outset that the model described in this section was derived for analysis of this specific dataset and its filtering/instrument background conditions, and may not be applicable to other observations and/or time periods. A full accounting of the updated NuSTAR background model will be the subject of upcoming work.
\par We begin by applying the original NuSTAR background model of Ref.~\cite{Wik:2014} to stacked data taken while NuSTAR's FOV was both occulted by the Earth (OCC, determined by the elevation angle ELV between the telescope boresight and Earth's limb) and shaded from the Sun (NOSUN, determined by an onboard sensor), during which the event rate was dominated by the internal detector background. (\textsc{nustardas} defines the OCC mode to begin when ELV$\,<3^\circ$; however, we find that this is not sufficient to fully suppress x-rays from the brightest sources near the limb of the Earth, so we require ELV$\,<0^\circ$.) After this filtering, the stacked OCC-mode spectra had exposures $\sim$17 Ms/detector, further reduced to $\sim$4.5 Ms/detector when the NOSUN filter was applied. (To ensure the Sun remains well below the horizon during NOSUN periods, we also exclude data 300 seconds before and after each period of solar illumination.) Principal component analysis showed no significant spatial variation in the internal background across the detectors. The internal detector background can be divided into two components: a featureless continuum and a large set of lines. The internal continuum model is the same as Ref.~\cite{Wik:2014}, i.e., a broken power law with $E_\text{break} = 121.86\text{\,keV}$ and spectral indices $\Gamma_1 = -0.047$ and $\Gamma_2 = -0.838$ for energies below and above $E_\text{break}$ respectively. (As before, we define the power law as $E_\gamma^{-\Gamma}$, i.e., the internal continuum increases with $E_\gamma$.) The internal continuum dominates the background for energies $E_\gamma \gtrsim 100\text{\,keV}$ and is taken to have the same shape (though potentially different normalizations) for each detector.
\par The internal detector background lines deserve special consideration, as narrow lines can mimic a DM signal. We begin by applying the list of Lorentzian lines (plus internal continuum) from Ref.~\cite{Wik:2014} to the data taken when the telescope's FOV was both occulted by the Earth and shielded from the Sun (OCC$\,\cap\,$NOSUN). Inspection of the residuals showed the need for additional wide lines to model the ``plateau'' observed in the continuum for energies ${\sim}$10--20\,keV. These features may result from the CXB-induced Earth x-ray albedo~\cite{Churazov:2008}, though further study is ongoing. To check for any drift in the line positions and/or widths over time, we further divide each detector's 2012--2017 OCC$\,\cap\,$NOSUN data into eight sequential periods of similar exposure time based on the observation date. \blue{As the detector chips share the same geometry and radiation environment, we expect their internal backgrounds to be highly correlated. Thus, we fit each stack's spectrum to the same internal continuum plus lines model described above. To account for uncorrelated variations between the chips (e.g., from un-modeled variations in detector gain), we also allow the line positions and widths to vary about their nominal positions}. The line positions and widths for the final model of each detector (shown in Table~\ref{table:parametric_model}) are then fixed to the weighted average of the best-fit values from each temporal stack. \blue{During our analysis, we allow the normalizations of the background lines to vary unless otherwise noted, to account for solar modulation, geomagnetic activity, and detector aging effects.}
\par With a working model of the internal background for each detector, we next consider the full OCC-mode spectra, including both SUN and NOSUN periods. A component following the 0-bounce spatial gradient is clearly visible in detector images at low energies $E_\gamma \lesssim 10\text{\,keV}$, indicating that solar x-rays (likely reflected from the mast and optics bench) are striking the detectors. Furthermore, the intensity of this component appears to be correlated to solar activity. This ``solar'' component includes both direct and reflected solar x-rays, and features a steeply-falling continuum and several narrow lines. We model the continuum as a simple power law to approximate the high-energy tail of a thermal plasma with temperature ${\sim}$few million K \cite{Hannah:2016,Glesener:2017,Wright:2017,Kuhar:2018lao}. We attribute the lines to a combination of direct solar illumination and fluorescence from the telescope structural elements. Similarly to the treatment of the internal detector lines, we divide the full OCC-mode spectra for each detector into eight temporal slices, calculate best-fit solar spectral indices and line positions/widths for each, and average the values from each epoch to obtain the values in Table~\ref{table:parametric_model}. This solar model is more flexible than the \texttt{apec} model of Ref.~\cite{Wik:2014}, as decoupling the continuum shape and line fluxes allows us to better model the solar background over a large fraction of the solar activity cycle. We caution that the solar power law and line parameters were derived with the specific filtering conditions used in this analysis, and will likely vary considerably (and unpredictably) with different solar cycle conditions.
\par Finally, we consider the cosmic x-ray background (CXB), which is the dominant astrophysical background once bright sources have been removed. As discussed previously, there are simultaneous 0-bounce and 2-bounce CXB contributions with the same underlying sky intensity ${\mathcal{I}_\text{CXB} \equiv d^2 F_\gamma/dE_\gamma d\Omega}$. The 2-bounce CXB is modulated by energy-dependent effective area $A_\text{2b}(E_\gamma)$ of the optics, whereas the 0-bounce CXB is modulated only by the geometry of the aperture stops, optics bench, and detectors. (Both CXB components are modulated by the Be window and detector dead layer throughputs $\mathcal{E}_\text{Be}$ and $\mathcal{E}_\text{det}$.) We adopt the same CXB model as Sec.~\ref{sec:roman_model}, though here we also include an interstellar medium absorption component via the \textsc{xspec} model \texttt{tbabs} with equivalent hydrogen column ${N_\text{H} = 4.7\times 10^{20}\text{\,cm}^{-2}}$ (averaging over all observations) and Solar elemental abundances \cite{Wilms:2000ez,Dickey:1990mf,Kalberla:2005ts,Anders:1989zg}. \blue{(Even at the lowest energy $E_\gamma = 3$ keV, the equivalent optical depth $\tau$ is still $\lesssim \text{few}\times 10^{-3}$. Though the expected attenuation is $\lesssim0.5\%$, we include it for completeness.)}
As the CXB spectral model and 0-bounce instrument response are well constrained at the ${\sim}$percent level~\cite{Krivonos:2020qvl}, we completely freeze the 0-bounce CXB component; however, we allow the 2-bounce CXB normalization a nominal $\pm 10\%$ range to account for residual uncertainties in the 2-bounce effective area and solid angle following point-source removal.

\begin{table}
    \centering
    \begin{tabularx}{\columnwidth}{XXXX}
        \hline
        \hline
        Detector & Exposure\,(Ms) & Avg.$A_\text{0b}$\,(cm$^2$) & Avg.$\Delta \Omega_\text{0b}$\,(deg$^2$)\\
        \hline
        A0 & 18.8 & 1.22 (3.18) & 2.20 (2.31) \\
        A1 & 19.9 & 2.07 (3.14) & 2.95 (2.82)\\
        A2 & 20.0 & 2.57 (3.17) & 6.75 (6.63)\\
        A3 & 19.6 & 2.00 (3.13) & 6.62 (6.32) \\
        B0 & 18.9 & 1.24 (3.18) & 7.22 (6.98) \\
        B1 & 19.5 & 2.27 (3.15) & 4.62 (4.63) \\
        B2 & 19.8 & 2.57 (3.19) & 1.14 (1.28)\\
        B3 & 19.9 & 1.76 (3.12) & 5.47 (5.41)\\
        \hline
        \hline
    \end{tabularx}
    \caption{NuSTAR detector exposures and grasps for the parametric analysis described in Sec.~\ref{sec:parametric}. The average $A_\text{0b}$ and $\Delta\Omega_\text{0b}$ are the exposure-time-weighted averages over the individual obsIDs calculated using \textsc{nuskybgd}. The values in parentheses correspond to observations with no astrophysical source regions masked.}
    \label{table:parametric_exposure}
\end{table}

\subsection{\label{sec:parametric_fitting}Spectral Fitting}
As the backgrounds for each of the eight NuSTAR detectors are slightly different, we individually fit each chip's cleaned on-sky science-mode spectrum (to be contrasted with the Earth-occulted OCC spectra) to the model described in Sec.~\ref{sec:parametric_model} and Table~\ref{table:parametric_model}. For all model components except the internal continuum (whose energy response is purely diagonal), we use the \textsc{v}3 redistribution matrix files (RMFs) from \textsc{caldb} \textsc{v}20211020 \cite{Madsen:2021hjz}. All model components except the internal continuum and internal lines also include the Be window and CdZnTe dead layer transmission efficiencies $\mathcal{E}_\text{Be}$ and $\mathcal{E}_\text{det}$, also taken from \textsc{caldb} \textsc{v}20211020. For the 0-bounce CXB, we use the effective areas $A_\text{0b}$ and solid angles $\Delta \Omega_\text{0b}$ from Table~\ref{table:parametric_exposure}. For the 2-bounce CXB, we construct FOV-averaged 
$\langle A_\text{2b} \Delta \Omega_\text{2b}\rangle$ for each detector by averaging the effective areas and solid angles from the individual observations (see Sec.~\ref{sec:allsky_data}). The $A_\text{2b}$ were calculated before the updated \textsc{caldb} \textsc{v}20211020 became available, though as described in Sec.~\ref{sec:parametric_model} we allow the 2-bounce CXB component to vary in overall normalization by $\pm10\%$ to account for residual uncertainties in the effective area. In any case, the impact on our DM constraints is expected to be marginal compared to the ${\sim}$20\% DM profile uncertainties.

\par With ${\sim}$20 Ms exposure per spectrum, even with the finest possible binning (one bin per 40-eV NuSTAR channel), the statistical uncertainty is at the level of a few percent per bin. The small statistical uncertainties allow systematic deviations---especially in the vicinity of background lines---to become visible. This is expected, as the line centroids and FWHMs are known to drift over the years. We address these systematics in two ways. First, we assign a flat 2.5\% systematic uncertainty added in quadrature to the statistical uncertainty in each bin, sufficient to give $\chi^2/\text{dof} \sim 1$. Second, as discussed in Sec.~\ref{sec:constraints_parametric}, we power constrain our DM limits to mitigate the effects of downward fluctuations of data with respect to the model.
\par Owing to the complexity of the full parametric background model, it was not computationally feasible to simulate and model the many mock datasets needed for sensitivity estimates. Instead, we constructed one ``Asimov'' dataset \cite{Cowan:2010js} per SCI-mode spectrum, in which the event rates per energy bin were set equal to their best-fit values (including the 2.5\% systematic) using the models described in Secs.~\ref{sec:parametric_model} and \ref{sec:parametric_fitting}. These Asimov spectra were passed through the same modeling and DM-search procedure as the real data.
\begin{table*}[]
    \centering
    \begin{tabular}{llll}
        \hline
        \hline
        Component & \textsc{xspec} model & Response & Free parameters\\
        \hline
        CXB (0b) & \texttt{tbabs}*\texttt{powerlaw}*\texttt{highecut} & Detector RMF, $\mathcal{E}_\text{Be}$, $\mathcal{E}_\text{det}$, $\langle A_\text{0b} \Delta \Omega_\text{0b}\rangle$ & None \\
        CXB (2b) & \texttt{tbabs}*\texttt{powerlaw}*\texttt{highecut} & Detector RMF, $\mathcal{E}_\text{Be}$, $\mathcal{E}_\text{det}$, $\langle A_\text{2b} \Delta \Omega_\text{2b}\rangle$ & 0-bounce flux $\pm 10\%$\\
        Internal continuum & \texttt{bknpower} & Diagonal RMF & Normalization \\
        Internal lines & $\sum$\texttt{lorentz} & Detector RMF & Normalizations\\
        Solar & \texttt{powerlaw} + $\sum$\texttt{lorentz} & Detector RMF, $\mathcal{E}_\text{Be}$, $\mathcal{E}_\text{det}$ & Powerlaw and line norms. \\
        DM line (0b) & \texttt{tbabs}*\texttt{gaussian} & Detector RMF, $\mathcal{E}_\text{Be}$, $\mathcal{E}_\text{det}$, $\langle A_\text{0b} \Delta \Omega_\text{0b}\rangle$ & See Sec.~\ref{sec:constraints_parametric} \\
        DM line (2b) & \texttt{tbabs}*\texttt{gaussian} & Detector RMF, $\mathcal{E}_\text{Be}$, $\mathcal{E}_\text{det}$, $\langle A_\text{2b} \Delta \Omega_\text{2b}\rangle$ & See Sec.~\ref{sec:constraints_parametric} \\
        \hline
        \hline
    \end{tabular}
    \caption{Summary of \textsc{xspec} model components for the parametric background analysis of Sec.~\ref{sec:parametric}. The detector RMFs and $\mathcal{E}_\text{det}$ are taken from \textsc{caldb} \textsc{v}20211020.}
    \label{tab:my_label}
\end{table*}

\section{\label{sec:constraints}NuSTAR DM Search}

\par With the background models described in Secs.~\ref{sec:roman_model} and ~\ref{sec:parametric_model} we search our spectra for evidence of DM decay lines using the same general procedure as our previous NuSTAR analyses \cite{Perez:2016tcq,Ng:2019a,Roach:2019ctw}, which we summarize below.
\subsection{\label{sec:stat_formalism}Statistical Formalism}
For each trial DM mass $m_\chi$ in a given spectrum, we search for evidence of DM using the profile likelihood ratio \cite{Cowan:2010js,Algeri:2020a}. We take the likelihood $\mathcal{L}$ to be a function of the count rate $\boldsymbol{y}$, the DM signal strength $\mu$ (here, the decay rate $\Gamma$), and the background model parameters $\boldsymbol{\eta}$. The test statistic (TS) in favor of the DM hypothesis is given by
\begin{equation}
    \text{TS}(\hat{\mu}|m_\chi) = -2\ln\left[\frac{\max_{\mu,\boldsymbol{\eta}} \mathcal{L}(\boldsymbol{y}|\boldsymbol{\eta},\mu))}{\max_{\boldsymbol{\eta}}\mathcal{L}(\boldsymbol{y}|\boldsymbol{\eta},\mu=0))}\right],
\end{equation}
where $\hat{\mu} \ge 0$ is the best-fit positive signal strength. We also define an analogous quantity $q$ used for obtaining an upper limit on the decay rate:
\begin{equation}
    q(\mu|m_\chi) = -2\ln\left[\frac{\max_{\boldsymbol{\eta}} \mathcal{L}(\boldsymbol{y}|\boldsymbol{\eta},\mu))}{\max_{\mu,\boldsymbol{\eta}}\mathcal{L}(\boldsymbol{y}|\boldsymbol{\eta},\mu))}\right].
\end{equation}
For each trial mass $m_\chi$, we scan through a range of signal strengths $\mu > 0$, allowing the background model parameters $\boldsymbol{\eta}$ to find their best-fit values. In particular, by allowing the DM line to assume the full strength of any background lines, we obtain conservative limits on the DM flux in the vicinity of these lines. In the large-count limit, the log-likelihood ratio (and hence TS and $q$) reduces to $\Delta \chi^2$ for a single degree of freedom:
\begin{equation}
    \begin{dcases}
        \blue{\text{TS}(\hat{\mu}|m_\chi) \simeq \text{min}_\mu \left[\chi^2 (\mu|m_\chi) \right]- \chi^2(\mu = 0|m_\chi) } \\
        \blue{q(\mu|m_\chi) \simeq \chi^2(\mu|m_\chi) - \chi^2(\hat{\mu}|m_\chi)  }
    \end{dcases}
\end{equation}

The detection significance in Gaussian standard deviations for a DM line in a single spectrum is thus simply $\sqrt{\text{TS}}$. In the absence of detections above the $5\sigma$ threshold, we set one-sided 95\% upper limits to be the signal strength $\mu_{95}$ where ${q(\mu_{95}|m_\chi) = 2.71 )}$. For the Asimov datasets described in Sec.~\ref{sec:parametric_fitting}, the $\pm N\sigma$ containment bands around the median expected 95\% upper limit occur where $q$ increases from its minimum by $(1.64 \pm N)^2$ \cite{Cowan:2010js,Foster:2017hbq}. To incorporate constraints from multiple spectra $k$, \blue{we consider the object $\mathfrak{X}^2(\mu|m_\chi) = \sum_k \chi_k^2(\mu|m_\chi)$. The corresponding expressions for $\text{TS}_\text{joint}$ and $q_\text{joint}$ are}
\begin{equation}
    \begin{dcases}
    \text{TS}_\text{joint}(\hat{\mu}_\text{joint}|m_\chi) = \blue{\text{min}_\mu \left[\mathfrak{X}^2(\mu|m_\chi)\right] - \mathfrak{X}^2(0|m_\chi)  } \\
    q_\text{joint}(\mu|m_\chi) = \blue{ \mathfrak{X}^2(\mu|m_\chi) - \mathfrak{X}^2(\hat{\mu}_\text{joint}|m_\chi) }
    \end{dcases}
\end{equation}
where $\hat{\mu}_\text{joint}$ is the joint maximum-likelihood signal strength considering all spectra \blue{(i.e., the minimum of $\mathfrak{X}^2$)}. We use Eq. (1) to convert the limits on $dF_\text{DM}/d\Omega$ to limits on the decay rate $\Gamma$, where the effective DM column density $\overline{\langle d\mathcal{D}/d\Omega \rangle}$ is obtained by averaging over both the FOV and the fractional exposure time $\Delta t_k/T$ of each observation:
\begin{equation}
    \overline{\left \langle \frac{d\mathcal{D}}{d\Omega} \right \rangle} = \sum_k\left[ \frac{\Delta t}{T \Delta \Omega} \int_\text{FOV} \xi\,\text{d}\Omega \int_\text{LOS} \rho_\chi\,d\text{s}\right]_k.
\end{equation}

\begin{figure*}[htp]
  \centering
  \subfigure{\includegraphics[scale=0.9]{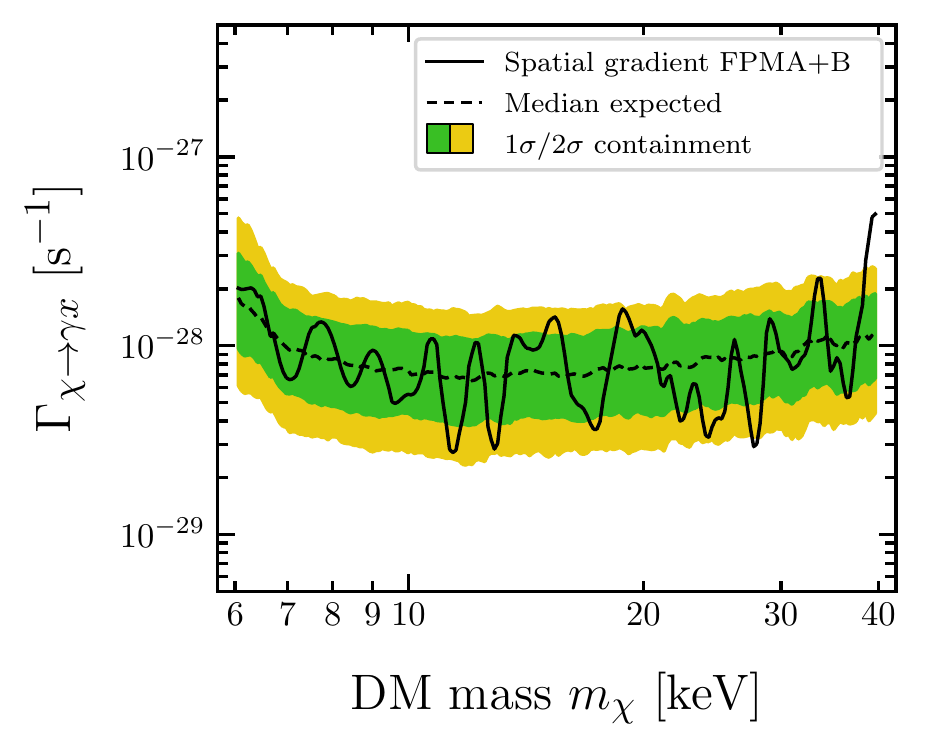}}
  \subfigure{\includegraphics[scale=0.9]{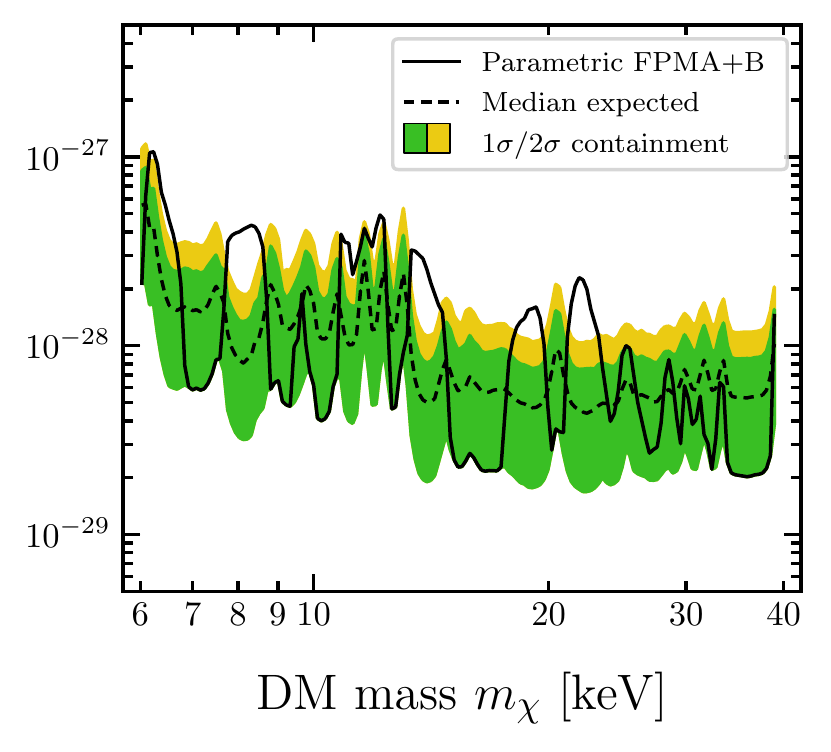}}
  
  \caption{\textbf{Left:} 95\% confidence upper limits and expected containment on the single-photon decay rate $\Gamma_{\chi\rightarrow \gamma x}$ for the spatial-gradient method of Sec.~\ref{sec:spatial}. \textbf{Right:} same quantity for the parametric method of Sec.~\ref{sec:parametric}. For two-photon decays $\chi\rightarrow \gamma \gamma$, e.g., for axionlike particles \cite{Arias:2012az,Irastorza:2018dyq,XENON:2020rca,Takahashi:2020bpq}, the decay-rate limits in both plots are strengthened by a factor of 2. A comparison of all NuSTAR limits may be found in Fig.~\ref{fig:nustar_all_limits}.}
  \label{fig:dmgamma_constraints}
\end{figure*}

\subsection{\label{sec:constraints_spatial}Constraints from Spatial-Gradient Analysis}
To the spectral models for FPMA and FPMB described in Sec.~\ref{sec:roman_model} we add a DM line convolved with the 0-bounce instrument response. We scan 200 DM masses uniformly spaced in $\log_{10} m_\chi$ between 6--40 keV for both the data and $10^3$ mock spectra (i.e., oversampling with respect to the detector energy resolution). The FPMA and FPMB spectra are scanned separately and the joint statistics are calculated as described previously. The obtained limits from both FPMA and B and their combination are in excellent agreement both with our simulations and the asymptotic expectations.  Aside from an upward fluctuation in the detection significance for masses ${\sim}$38--40 keV, resulting from a few upward-fluctuating bins in each spectrum, we find no evidence of x-ray lines, demonstrating the power of the spatial-gradient technique for suppressing detector backgrounds. We note that a DM interpretation for the aforementioned excess is inconsistent with previous NuSTAR constraints \cite{Neronov:2016wdd,Perez:2016tcq,Ng:2019a,Roach:2019ctw}. Unlike the parametric limits described in Sec.~\ref{sec:constraints_parametric}, we do not power constrain the spatial-gradient limits, as we are able to generate both $\pm 1\sigma$ and $\pm 2\sigma$ containment bands by bootstrapping from our $10^3$ simulated spectra.
\par The dominant systematic uncertainty on the spatial-gradient limits arises from the choice of DM profile. The NFW profile gives a column density ${\overline{\langle d\mathcal{D}/d\Omega \rangle} \approx 5 \text{\,GeV\,cm}^{-3}\text{\,kpc\,sr}^{-1}}$ at the position of these observations ($\sim 100^\circ$ from the Galactic Center) with the shallow (sNFW) profile giving a column density ${\sim}15\%$ higher. On the other hand, the profile proposed by Ref.~\cite{Cautun:2019eaf} gives a value ${\sim}25\%$ lower than our default NFW profile, a consequence of the contracted halo. Our spatial-gradient limits shown in Figs.~\ref{fig:dmgamma_constraints} and \ref{fig:vMSM} are derived using the NFW profile, which we take as a ``median'' column density.

\subsection{\label{sec:constraints_parametric}Constraints from Parametric-Modeling Analysis}
To search for evidence of decaying DM in the eight single-detector spectra (both data and Asimov mock spectra) of Sec.~\ref{sec:parametric_fitting}, we adopt the same scanning strategy as described in Sec.~\ref{sec:constraints_spatial}, with two key differences. First, the background model is substantially more complex, a consequence of the many activation/fluorescence lines. As we are searching for anomalous x-ray lines in the energy range 3--20 keV, we freeze the normalizations of all lines between 20--95 keV to their best-fit values under the null-DM hypothesis. This procedure ensures that the minimizer does not become stuck in irrelevant local minima and greatly increases the scanning speed, with negligible impacts on our DM constraints compared to tests in which all lines are free to fit. Additionally, unlike the spatial-gradient case in which the 2-bounce contribution to the DM signal was negligible, here we model both the 0-bounce and 2-bounce contributions. Using the NFW profile, we find $\overline{\langle d\mathcal{D}/d\Omega} \rangle\approx 5\text{\,GeV\,cm}^{-3}\text{\,kpc\,sr}^{-1}$ for the 0-bounce and 2-bounce apertures, though this may vary in either direction by ${\sim}$20--25\% if the sNFW or contracted profiles are considered. Scanning the eight individual-detector spectra $k$ with a grid of 165 masses between 6--40 keV evenly spaced in $\log_{10} m_\chi$, we collect the distributions $q_k(\mu|m_\chi)$. (This is somewhat smaller than 200 mass bins in the spatial-gradient analysis of Sec.~\ref{sec:spatial}, owing to the much greater complexity and computational cost of the full parametric model.)
\par Similarly to Sec.~\ref{sec:constraints_spatial}, we calculate the line detection significance and one-sided 95\% confidence upper limits for each of the eight spectra individually, as well as the joint constraints summing over all eight spectra. We identify five mass ranges in which the joint decay-rate limit significantly worsens compared to the expected values from the Asimov procedure: (i) ${\sim}$7.8--8.5 keV, (ii) ${\sim}$11--12 keV, (iii) ${\sim}$13--14 keV, (iv) ${\sim}$18--19 keV, and (v) ${\sim}$21--22 keV. In particular, the excesses in (i), (iii), and (v) have local significance $>5\sigma$, and are observed on multiple detectors on both FPMs. We argue that none of these five excesses are consistent with decaying DM for several reasons. First, the spatial-gradient analysis finds no excesses with significance $>2\sigma$ in these mass ranges, strongly constraining any astrophysical origin. Second, we note that all of these excesses occur near adjacent instrumental and/or solar lines, causing any mismodeling of these lines or the instrument response to be amplified. Finally, we observe that the excesses do not have consistent fluxes $dF/d\Omega$ across the detectors, contrary to the expectation of an approximately uniform intensity from decaying DM across the instrument FOV. [In particular, we note the strong excess at $E_\gamma \sim 4 \text{\,keV}$ ($m_\chi \sim 8 \text{\,keV}$) on detector B2, despite this detector having a very small solid angle $\Delta \Omega_\text{0b}$.]
\par In the absence of plausible DM detections, we instead set conservative upper limits on the decay rate, which would not exclude a DM signal with these decay rates if it were present. To avoid setting artificially strong limits as a result of downward fluctuations, and because the simple Asimov procedure we employed cannot be used to define the $-2\sigma$ edge of the 95\% confidence band, we power constrain our limits \cite{Cowan:2011an}, i.e., the observed limit cannot run below the $(\text{median}-1\sigma)$ level expected from the Asimov simulations. These power-constrained limits are shown in Fig.~\ref{fig:dmgamma_constraints}.

\begin{figure}[b!]
    \includegraphics[scale=0.95,left]{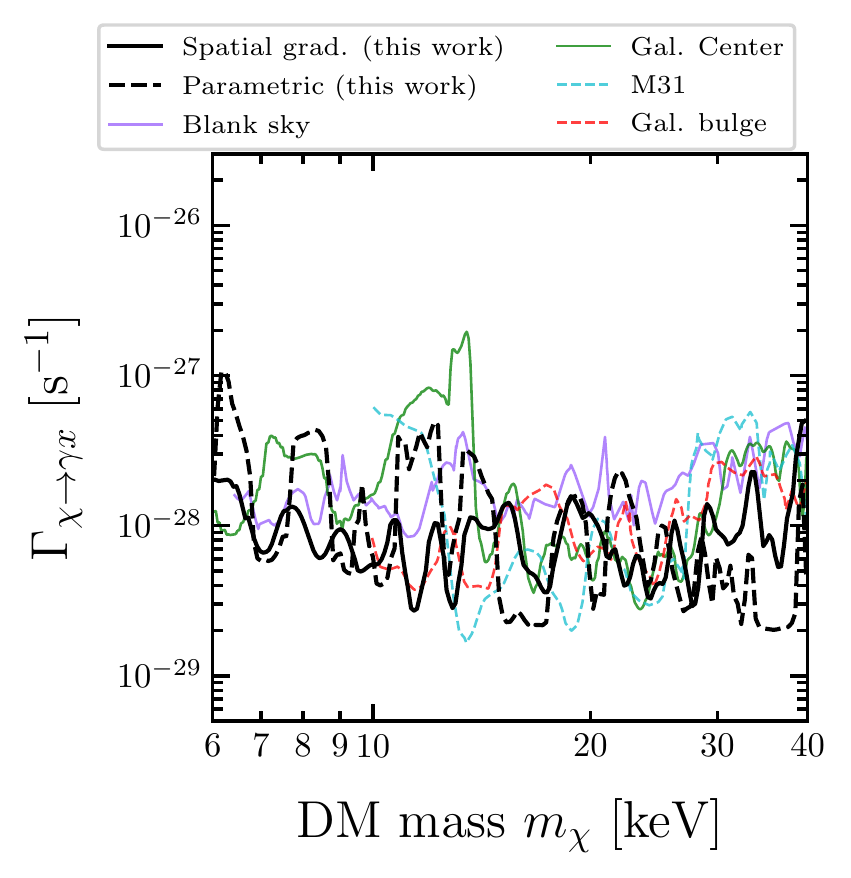}
    \caption{Comparison of NuSTAR DM decay-rate limits from this work (black solid and dashed lines) and previous NuSTAR analyses of blank-sky extragalactic fields~\cite{Neronov:2015kca}, the Galactic Center~\cite{Perez:2016tcq}, M31~\cite{Ng:2019a}, and the Galactic bulge~\cite{Roach:2019ctw}.}
    \label{fig:nustar_all_limits}
\end{figure}

\begin{figure*}[htp]
  \centering
  \subfigure{\includegraphics[scale=1.05]{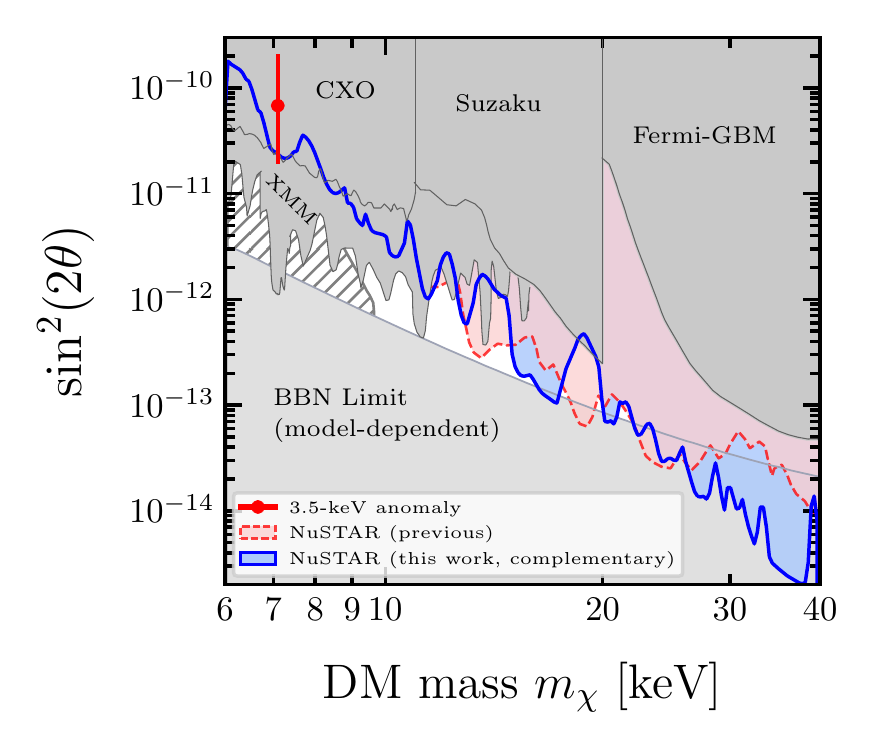}}
  \hspace*{-0.5cm}
  \subfigure{\includegraphics[scale=1.05]{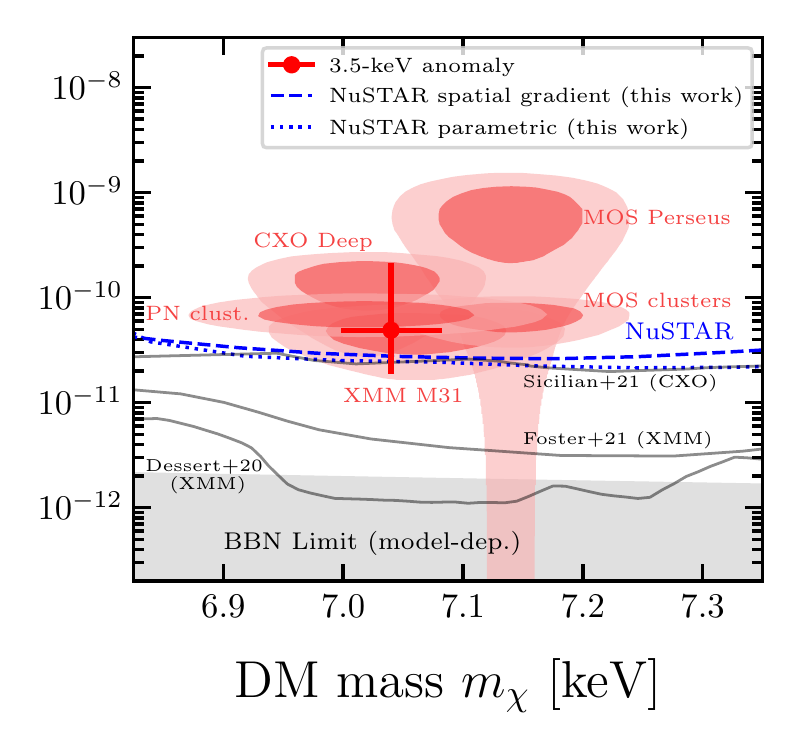}}
  
  \caption{Constraints on sterile-neutrino DM in the $\nu\text{MSM}$. \textbf{Left:} Constraints on $\sin^2(2\theta)$ from previous NuSTAR analyses (light red shade and solid line, Refs.~\cite{Perez:2016tcq,Neronov:2016wdd,Ng:2019a,Roach:2019ctw}) and other x-ray telescopes (dark gray, Refs.~\cite{Dessert:2018qih,Foster:2021ngm,Sicilian:2020a,Tamura:2014mta,Ng:2015gfa}) compared to the complementary NuSTAR constraints from this work (dark blue shade and solid line showing the parameter space excluded by either the spatial-gradient or parametric analysis, whichever is stronger at each mass) and an example of the 3.5-keV anomaly including DM profile uncertainties (red point, Ref.~\cite{Boyarsky:2014jta}). The gray hatched region represents structure-formation constraints from the observed number of MW dwarf satellites \cite{Cherry:2017dwu}. \textbf{Right:} Enlarged version of the previous plot. The dark (light) red contours show the $1\sigma$ ($2\sigma$) detection regions of the 3.5-keV anomaly as detailed in Fig.~7 of Ref.~\cite{Abazajian:2017tcc}. The red cross representing the 3.5-keV anomaly is the same as the left-hand figure. Recent deep constraints from CXO~($3\sigma$ upper limit, Ref.~\cite{Sicilian:2020a}) and XMM-Newton~(95\% upper limit, Refs.~\cite{Dessert:2018qih,Foster:2021ngm}) are compared to the NuSTAR constraints from this work (95\% upper limit, blue dashed/dotted lines). See Sec.~\ref{sec:sterile_constraints} for further details.}
  \label{fig:vMSM}
\end{figure*}

\subsection{\label{sec:sterile_constraints}Sterile-Neutrino DM Constraints}

\par Specializing to the particular case of sterile-neutrino decays $\chi \rightarrow \gamma\nu$, we convert our limits on the model-independent single-photon decay rate $\Gamma_{\chi\rightarrow \gamma x}$ to corresponding limits on the active-sterile mixing angle $\theta$ for Majorana neutrinos \cite{Shrock:1974nd,Pal:1981rm}:
\begin{equation}
    \Gamma_{\chi\rightarrow\gamma\nu} = 1.38\times 10^{-32} \text{\,s}^{-1} \left[\frac{\sin^2(2\theta)}{10^{-10}}\right] \bigg(\frac{m_\chi}{1\text{\,keV}}\bigg)^5. 
\end{equation}
Fig.~\ref{fig:vMSM} shows the upper limits on the sterile-neutrino DM parameter space obtained in this work.  For comparison, we also show previous limits from other x-ray searches from CXO \cite{Sicilian:2020a}, XMM-Newton \cite{Dessert:2018qih,Foster:2021ngm}, Suzaku \cite{Tamura:2014mta}, Fermi-GBM \cite{Ng:2015gfa}, and INTEGRAL-SPI \cite{Boyarsky:2007ge}. Our results further constrain the sterile-neutrino decay rate, especially for masses ${\sim}$15--18 keV and ${\sim}$25--40 keV, where the limit improves on previous NuSTAR constraints by a factor ${\sim}$2--3 and ${\sim}$5--10, respectively compared to previous NuSTAR constraints. We emphasize that these results are not specific to sterile-neutrino DM, but are also applicable to generic decaying DM models that involve a photon line, with the decay-rate limits given in Fig.~\ref{fig:dmgamma_constraints}. 
\par With improved modeling in the low-energy NuSTAR background, our limits extend down to DM masses ${m_\chi = 6\text{\,keV}}$, and have improved the limit by nearly an order of magnitude compared to previous NuSTAR results for masses below 10 keV~\cite{Perez:2016tcq, Neronov:2016wdd}. Importantly, our results are now in tension with the claimed tentative signal at $E_\gamma \simeq 3.5$\,keV~\cite{Bulbul:2014sua, Boyarsky:2014jta}.  Previous NuSTAR analyses included a line at 3.5 keV \cite{Perez:2016tcq,Neronov:2016wdd}, attributed to instrument background due to its presence in Earth-occulted data, though a possible astrophysical contribution was debated. Our present results constrain the DM origin of the 3.5-keV line in complementary ways. First, the spatial-gradient analysis of Sec.~\ref{sec:spatial} does not detect the 3.5-keV line. As the spatial-gradient technique suppresses detector backgrounds while remaining sensitive to astrophysical emission, its nonobservation simultaneously strongly constrains its astrophysical origin and favors its detector origin. Using the more traditional parametric modeling approach on a disjoint dataset covering a much larger area of the Galactic halo, we apply the improved model of the NuSTAR instrument background in Sec.~\ref{sec:parametric_model}. In particular, this background model includes a $\sim$3.5-keV line detected in Earth-occulted data when no DM events are expected. The best-fit width of this line ($\sim$0.5-keV FWHM prior to convolution with the detector response) is notably wider than the ${\sim}$0.4-keV FWHM detector resolution expected for an astrophysical line, suggesting it may be an artifact of variations in the detector dead layer absorption or instrument response (which are rapidly varying at low energies) rather than a genuine spectral line. Whatever its origin, we are still able to set conservative constraints on a 3.5-keV astrophysical line by allowing the DM line in our search procedure to incorporate events from the internal background. 
\par Modeling the spectra from each NuSTAR detector independently for the first time and combining the constraints, we obtain similar constraints to the spatial-gradient method across much of the 6--40 keV mass range, demonstrating the complementarity and consistency of the two approaches. Our null results on the 3.5\,keV line obtained with standard statistical techniques are in agreement with recent results from CXO \cite{Sicilian:2020a} and XMM-Newton \cite{Dessert:2018qih,Bhargava:2020a,Foster:2021ngm} despite NuSTAR having a substantially lower effective area and shorter exposure time than either, illustrating the power of NuSTAR's wide FOV. (We note that the limits in Ref.~\cite{Dessert:2018qih} have led to much discussion in the literature~\cite{Abazajian:2020unr,Boyarsky:2020hqb,Dessert:2020hro}.) Taken together, our results therefore provide strong and independent evidence against the astrophysical and DM interpretation of the 3.5-keV line in the Galactic halo.
%
\par In addition to the strong constraints at low masses, our present NuSTAR results are the strongest x-ray constraints on sterile-neutrino DM to date in the mass ranges ${\sim}$15--18 keV and ${\sim}$25--40 keV, improving on previous NuSTAR work by a factor ${\sim}$2--3 and ${\sim}$5--10, respectively (see Fig.~\ref{fig:nustar_all_limits}). In particular, improvements in the parametric background model in the energy range ${\sim}$15--20\,keV (mass range ${\sim}$30--40\,keV) allow us to leverage the statistical power of the full ${\sim}$20-Ms dataset without being limited by modeling systematics as in previous works (see, e.g., Refs.~\cite{Ng:2019a,Roach:2019ctw}). These new results further demonstrate NuSTAR's ability to provide DM constraints in this challenging mass range, as the use of 0-bounce photons allows the observatory to sample large areas of the sky without being limited by the strong energy dependence of x-ray mirrors. Finally, we note that the NuSTAR limits in this work are weaker than previous NuSTAR constraints from M31~\cite{Ng:2019a} and the Galactic bulge~\cite{Roach:2019ctw} near $m_\chi \sim 14 \text{\,keV}$ and $m_\chi \sim 19\text{\,keV}$. At 14 keV, the FPMB spatial-gradient spectrum experiences a mild excess at $E_\gamma \sim 7\text{\,keV}$, and the sensitivity of the parametric-modeling method is limited by a bright instrumental line at the same energy. At 19 keV, a similar weak excess is observed in the FPMB spatial-gradient spectrum, and the parametric model has difficulty reproducing the shape of the instrumental line at $E_\gamma \sim 10 \text{\,keV}$.

\par While our results are applicable to generic sterile-neutrino DM that mixes with SM neutrinos, they also have important implications for particular realizations of sterile-neutrino DM models.  One popular scenario is sterile-neutrino DM produced by mixing with active neutrinos~\cite{Dodelson:1993je, Shi:1998km}. If more right-handed neutrinos are also present, e.g., in the $\nu$MSM~\cite{Asaka:2005an, Asaka:2006nq, Canetti:2012vf,Canetti:2012kh}, it is also possible to explain baryogenesis and the origin of neutrino mass, solving three important problems in fundamental physics in the same framework. 
\par Assuming that the DM is resonantly produced in the presence of a primordial lepton asymmetry~\cite{Venumadhav:2015pla}, big bang nucleosynthesis (BBN) constraints on the lepton asymmetry~\cite{Serpico:2005bc} can therefore be used to set lower limits on the mixing angle $\theta$, below which resonant production would underproduce DM compared to its observed abundance. In Fig.~\ref{fig:vMSM} we show the BBN constraints from \textsc{sterile-dm}~\cite{Venumadhav:2015pla}, adopting the lepton asymmetry per unit entropy density $L_6 \equiv 10^6(n_\nu - n_{\bar{\nu}})/s \leq 2500$. Additional discussion of these BBN limits may be found in, e.g., Refs.~\cite{Laine:2008pg,Boyarsky:2009ix,Ghiglieri:2015jua,Cherry:2017dwu,Boyarsky:2018tvu,Roach:2019ctw}.
\par The velocity distribution of DM can suppress the formation of small-scale cosmological structure; thus, observational probes such as dwarf MW satellite galaxies~\cite{Schneider:2016uqi,Cherry:2017dwu,Dekker:2021scf} and the Lyman-$\alpha$ forest~\cite{Baur:2017stq,Yeche:2017upn,Garzilli:2018jqh,Palanque-Delabrouille:2019iyz} can be used to placed limits on the ``warmness'' of sterile neutrino DM, thereby constraining sterile-neutrino mixing parameters. In Fig.~\ref{fig:vMSM}, we show the MW satellite limit from Ref.~\cite{Cherry:2017dwu} (consistent with that from~\cite{Dekker:2021scf}).  We note, however, that these limits depend on both the sterile neutrino production physics \cite{Ghiglieri:2015jua,Venumadhav:2015pla} as well as the complex structure-formation processes needed to connect DM halos to the observed satellite galaxies. These structure-formation processes have substantial uncertainties, and stronger/weaker constraints (see, e.g., Refs.~\cite{Schneider:2014rda,DES:2020fxi,Newton:2020cog}) can be obtained with different models of galaxy formation~\cite{Dekker:2021scf}. Even if only the more conservative structure-formation limits of Ref.~\cite{Cherry:2017dwu} are considered, it is clear that the combination of several types of constraints have nearly closed the window for keV-range sterile neutrino dark matter, and that new work is needed to ensure robust sensitivity across the entire range.

\section{\label{sec:conclusions}Conclusions}
In this work, we obtain updated limits on DM decaying into monoenergetic photons with the NuSTAR x-ray observatory. We consider two complementary analyses conducted on disjoint datasets to leverage the full power of the available NuSTAR data: the spatial-gradient method, utilizing a novel geometric technique to greatly suppress the detector background; and a more traditional parametric method, combining a large amount of data with an updated model of the NuSTAR instrument background. Significantly, we are able to use the full NuSTAR energy range down to ${E_\gamma = 3\text{\,keV}}$, allowing us to sensitively test lower-mass DM candidates. These analyses complement and extend previous NuSTAR DM searches in the Milky Way and the M31 galaxy, which have large amounts of DM \cite{Neronov:2016wdd,Perez:2016tcq,Ng:2019a,Roach:2019ctw}.

\par Our new analyses provide significant DM constraints in two key mass ranges. First, our improved treatment of the low-energy instrument background allows us to strongly constrain a possible DM origin of the 3.5-keV anomaly using standard statistical techniques. Second, our constraints on DM masses ${\sim}$15--40 keV continue to fill in the sterile-neutrino parameter space down to---and below---the BBN limit. Our results are also applicable to other DM candidates decaying or annihilating into monoenergetic photons, e.g., axionlike particles \cite{Arias:2012az,Irastorza:2018dyq,XENON:2020rca,Takahashi:2020bpq}. If taking the latest results from satellite counting into account~\cite{DES:2020fxi,Dekker:2021scf}, the full parameter space is now mostly covered, which is an important milestone.  While this does not fully rule out sterile neutrinos as the only DM component, it does show that the simple and elegant mixing production mechanism~\cite{Dodelson:1993je, Shi:1998km} may be insufficient, and more involved modeling \cite{Kusenko:2006rh,Shaposhnikov:2006xi,Adulpravitchai:2014xna,Drewes:2015eoa} may be required to make sterile neutrinos a viable DM candidate. Ongoing and near-term missions such as Spektr-RG \cite{eROSITA:2020emt,Pavlinsky:2021ynp}, Micro-X \cite{Hubbard:2020bfl}, and XRISM \cite{XRISMScienceTeam:2020rvx}, and proposed missions such as Athena~\cite{Nandra:2013shg}, AXIS~\cite{Mushotzky:2018wio}, eXTP~\cite{eXTP:2016rzs}, {HEX-P}~\cite{Madsen:2019hex}, and Lynx~\cite{LynxTeam:2018usc} are hoped to further constrain the sterile-neutrino parameter space using various detector architectures and observing strategies \cite{Neronov:2015kca,Speckhard:2015eva,Caputo:2019djj,Lovell:2019aol,Zhong:2020wre,Dekker:2021bos}.

\section*{Acknowledgements}
We are grateful to the NuSTAR team for the excellent performance of the observatory and their assistance with data processing. We also thank Alexey Boyarsky, Josh Foster, Nick Rodd, Field Rogers, Mengjiao Xiao, and the anonymous reviewer for helpful comments and discussions. B.M.R. and K.P. thank Paul Acosta, Gabriel Collin, and the MIT Laboratory for Nuclear Science for computing support. B.M.R. and K.P. were supported by the Cottrell Scholar Award, Research Corporation for Science Advancement (RCSA), ID No. 25928. S.R. and D.R.W. were supported by NASA grant No. 80NSSC18K0686. K.C.Y.N. was supported by the RGC of HKSAR, project No. 24302721. J.F.B. was supported by U.S. National Science Foundation (NSF) grant No. PHY-2012955. B.W.G. was supported by NASA contract No.~NNG08FD60C. S.H. was supported by the U.S. Department of Energy Office of Science under award No. DE-SC0020262, NSF grants No.~AST-1908960 and No.~PHY-1914409, JSPS KAKENHI grant No.~JP22K03630, and the World Premier International Research Center Initiative (WPI Initiative), MEXT, Japan. R.K. was supported by Russian Science Foundation grant No.~22-12-00271. This research has made use of data and/or software provided by the High Energy Astrophysics Science Archive Research Center (HEASARC), which is a service of the Astrophysics Science Division at NASA/GSFC.

%


\onecolumngrid
\clearpage

\appendix

\section{\label{sec:appA}{Parametric Background Model}}
In this section, we list the values of the parameters used to define the background model of Sec. IV. The values of these
parameters are all frozen during our dark-matter analysis.
\setcounter{table}{0}
\renewcommand{\thetable}{\Alph{section}\arabic{table}}

\begin{table*}[h!]
\footnotesize
\begin{tabular}{c|l|llllllll}
\hline
\hline
\multicolumn{1}{l|}{}                   &                            & \multicolumn{4}{c|}{FPMA}                                                                                                                       & \multicolumn{4}{c}{FPMB}                                                                                                 \\ \hline
\multicolumn{1}{c|}{Model}    & \multicolumn{1}{c|}{Param.}           & \multicolumn{1}{c|}{Det. A0}        & \multicolumn{1}{c|}{Det. A1}         & \multicolumn{1}{c|}{Det. A2}        & \multicolumn{1}{c|}{Det. A3}         & \multicolumn{1}{c|}{Det. B0}        & \multicolumn{1}{c|}{Det. B1}        & \multicolumn{1}{c|}{Det. B2}        & 
\multicolumn{1}{c}{Det. B3}        \\ \hline
\multirow{44}{*}{Int. lines}        &    \multirow{44}{*}{$E_0$, FWHM}         & \multicolumn{1}{l|}{3.47, 0.76}   & \multicolumn{1}{l|}{3.53, 0.81}    & \multicolumn{1}{l|}{3.57, 0.82}   & \multicolumn{1}{l|}{3.50, 0.68}    & \multicolumn{1}{l|}{3.65, 0.84}   & \multicolumn{1}{l|}{3.62, 0.83}   & \multicolumn{1}{l|}{3.47, 0.67}   & 3.43, 1.22   \\
                                        &              & \multicolumn{1}{l|}{4.44, 0.00}   & \multicolumn{1}{l|}{4.40, 0.05}    & \multicolumn{1}{l|}{4.38, 0.26}   & \multicolumn{1}{l|}{4.30, 0.22}    & \multicolumn{1}{l|}{4.86, 0.01}   & \multicolumn{1}{l|}{4.40, 0.00}   & \multicolumn{1}{l|}{4.46, 0.15}   & 4.57, 0.14   \\
                                        &              & \multicolumn{1}{l|}{5.12, 1.43}   & \multicolumn{1}{l|}{5.16, 1.41}    & \multicolumn{1}{l|}{5.08, 1.21}   & \multicolumn{1}{l|}{5.01, 0.87}    & \multicolumn{1}{l|}{5.15, 1.37}   & \multicolumn{1}{l|}{5.17, 1.59}   & \multicolumn{1}{l|}{4.89, 0.00}   & 4.89, 0.12   \\
                                        &              & \multicolumn{1}{l|}{6.20, 0.00}   & \multicolumn{1}{l|}{6.19, 0.02}    & \multicolumn{1}{l|}{6.12, 0.00}   & \multicolumn{1}{l|}{6.13, 0.00}    & \multicolumn{1}{l|}{6.19, 0.09}   & \multicolumn{1}{l|}{6.11, 0.14}   & \multicolumn{1}{l|}{6.50, 0.00}   & 6.13, 0.00   \\
                                        &                            & \multicolumn{1}{l|}{7.40, 0.00}   & \multicolumn{1}{l|}{7.39, 0.05}    & \multicolumn{1}{l|}{7.38, 0.12}   & \multicolumn{1}{l|}{7.34, 0.02}    & \multicolumn{1}{l|}{7.40, 0.00}   & \multicolumn{1}{l|}{7.42, 0.05}   & \multicolumn{1}{l|}{7.90, 0.00}   & 7.43, 0.00   \\
                                        &                            & \multicolumn{1}{l|}{7.90, 0.00}   & \multicolumn{1}{l|}{7.90, 0.00}    & \multicolumn{1}{l|}{7.90, 0.00}   & \multicolumn{1}{l|}{7.84, 0.00}    & \multicolumn{1}{l|}{7.90, 0.00}   & \multicolumn{1}{l|}{7.90, 0.00}   & \multicolumn{1}{l|}{8.05, 2.51}   & 7.86, 2.78   \\
                                        &                            & \multicolumn{1}{l|}{8.17, 3.30}   & \multicolumn{1}{l|}{8.15, 3.15}    & \multicolumn{1}{l|}{7.93, 2.89}   & \multicolumn{1}{l|}{7.88, 2.81}    & \multicolumn{1}{l|}{7.97, 3.06}   & \multicolumn{1}{l|}{8.20, 3.56}   & \multicolumn{1}{l|}{8.74, 0.00}   & 7.88, 0.08   \\
                                        &                            & \multicolumn{1}{l|}{8.66, 0.06}   & \multicolumn{1}{l|}{8.64, 0.00}    & \multicolumn{1}{l|}{8.64, 0.00}   & \multicolumn{1}{l|}{8.64, 0.01}    & \multicolumn{1}{l|}{8.75, 0.00}   & \multicolumn{1}{l|}{8.75, 0.10}   & \multicolumn{1}{l|}{9.14, 0.42}   & 8.75, 0.17   \\
                                        &                            & \multicolumn{1}{l|}{10.40, 0.58}  & \multicolumn{1}{l|}{10.24, 0.34}   & \multicolumn{1}{l|}{10.31, 0.37}  & \multicolumn{1}{l|}{10.19, 0.69}   & \multicolumn{1}{l|}{10.18, 0.50}  & \multicolumn{1}{l|}{10.29, 0.44}  & \multicolumn{1}{l|}{10.27, 0.73}  & 10.26, 0.88  \\
                                        &                            & \multicolumn{1}{l|}{11.40, 19.37} & \multicolumn{1}{l|}{11.45, 16.85}  & \multicolumn{1}{l|}{11.04, 19.28} & \multicolumn{1}{l|}{11.69, 19.37}  & \multicolumn{1}{l|}{11.63, 19.36} & \multicolumn{1}{l|}{11.02, 18.34} & \multicolumn{1}{l|}{10.79, 19.07} & 11.69, 0.00  \\
                                        &                            & \multicolumn{1}{l|}{12.60, 0.00}  & \multicolumn{1}{l|}{12.60, 0.00}   & \multicolumn{1}{l|}{12.60, 0.00}  & \multicolumn{1}{l|}{12.60, 0.00}   & \multicolumn{1}{l|}{12.60, 0.00}  & \multicolumn{1}{l|}{12.60, 0.00}  & \multicolumn{1}{l|}{12.06, 0.00}  & 12.60, 0.00  \\
                                        &                            & \multicolumn{1}{l|}{13.20, 19.37} & \multicolumn{1}{l|}{13.17, 0.77}   & \multicolumn{1}{l|}{13.17, 0.00}  & \multicolumn{1}{l|}{12.74, 18.47}  & \multicolumn{1}{l|}{13.20, 20.00} & \multicolumn{1}{l|}{13.11, 20.00} & \multicolumn{1}{l|}{13.17, 0.00}  & 12.76, 19.37 \\
                                        &                            & \multicolumn{1}{l|}{13.80, 19.37} & \multicolumn{1}{l|}{13.20, 19.37}  & \multicolumn{1}{l|}{13.20, 19.37} & \multicolumn{1}{l|}{13.17, 0.00}   & \multicolumn{1}{l|}{13.80, 20.00} & \multicolumn{1}{l|}{13.17, 0.76}  & \multicolumn{1}{l|}{13.20, 19.37} & 13.23, 0.00  \\
                                        &                            & \multicolumn{1}{l|}{14.20, 0.00}  & \multicolumn{1}{l|}{13.80, 19.37}  & \multicolumn{1}{l|}{13.80, 0.00}  & \multicolumn{1}{l|}{13.80, 0.00}   & \multicolumn{1}{l|}{14.20, 0.00}  & \multicolumn{1}{l|}{13.80, 0.00}  & \multicolumn{1}{l|}{13.93, 0.08}  & 14.17, 1.13  \\
                                        &                            & \multicolumn{1}{l|}{15.05, 0.38}  & \multicolumn{1}{l|}{14.99, 0.50}   & \multicolumn{1}{l|}{15.01, 0.27}  & \multicolumn{1}{l|}{14.84, 0.08}   & \multicolumn{1}{l|}{14.92, 0.00}  & \multicolumn{1}{l|}{14.71, 1.13}  & \multicolumn{1}{l|}{14.74, 0.73}  & 14.83, 0.28  \\
                                        &                            & \multicolumn{1}{l|}{16.00, 0.00}  & \multicolumn{1}{l|}{15.55, 0.05}   & \multicolumn{1}{l|}{15.95, 0.00}  & \multicolumn{1}{l|}{15.64, 0.00}   & \multicolumn{1}{l|}{15.73, 0.32}  & \multicolumn{1}{l|}{15.71, 0.09}  & \multicolumn{1}{l|}{15.72, 0.00}  & 15.80, 0.00  \\
                                        &                            & \multicolumn{1}{l|}{16.72, 0.10}  & \multicolumn{1}{l|}{16.63, 0.07}   & \multicolumn{1}{l|}{16.64, 0.00}  & \multicolumn{1}{l|}{16.47, 0.00}   & \multicolumn{1}{l|}{16.79, 0.00}  & \multicolumn{1}{l|}{16.64, 0.00}  & \multicolumn{1}{l|}{16.67, 0.06}  & 16.79, 0.00  \\
                                        &                            & \multicolumn{1}{l|}{19.56, 0.00}  & \multicolumn{1}{l|}{19.30, 0.10}   & \multicolumn{1}{l|}{19.55, 0.58}  & \multicolumn{1}{l|}{19.49, 0.58}   & \multicolumn{1}{l|}{19.60, 0.25}  & \multicolumn{1}{l|}{19.54, 0.05}  & \multicolumn{1}{l|}{19.52, 0.13}  & 19.52, 0.01  \\
                                        &                            & \multicolumn{1}{l|}{21.82, 0.63}  & \multicolumn{1}{l|}{21.81, 0.66}   & \multicolumn{1}{l|}{21.85, 0.98}  & \multicolumn{1}{l|}{21.81, 1.42}   & \multicolumn{1}{l|}{21.82, 0.59}  & \multicolumn{1}{l|}{21.75, 0.65}  & \multicolumn{1}{l|}{21.75, 0.68}  & 21.75, 0.48  \\
                                        &                            & \multicolumn{1}{l|}{22.86, 0.01}  & \multicolumn{1}{l|}{22.85, 0.06}   & \multicolumn{1}{l|}{22.89, 0.27}  & \multicolumn{1}{l|}{22.86, 0.57}   & \multicolumn{1}{l|}{22.92, 0.00}  & \multicolumn{1}{l|}{22.89, 0.11}  & \multicolumn{1}{l|}{22.89, 0.19}  & 22.93, 0.18  \\
                                        &                            & \multicolumn{1}{l|}{24.71, 1.65}  & \multicolumn{1}{l|}{24.74, 1.69}   & \multicolumn{1}{l|}{24.71, 1.52}  & \multicolumn{1}{l|}{24.66, 1.56}   & \multicolumn{1}{l|}{24.72, 1.82}  & \multicolumn{1}{l|}{24.76, 1.82}  & \multicolumn{1}{l|}{24.72, 1.70}  & 24.83, 2.02  \\
                                        &                            & \multicolumn{1}{l|}{25.23, 0.14}  & \multicolumn{1}{l|}{25.22, 0.16}   & \multicolumn{1}{l|}{25.23, 0.18}  & \multicolumn{1}{l|}{25.19, 0.18}   & \multicolumn{1}{l|}{25.23, 0.17}  & \multicolumn{1}{l|}{25.22, 0.15}  & \multicolumn{1}{l|}{25.23, 0.21}  & 25.24, 0.20  \\
                                        &                            & \multicolumn{1}{l|}{27.90, 1.21}  & \multicolumn{1}{l|}{27.92, 1.39}   & \multicolumn{1}{l|}{27.97, 0.00}  & \multicolumn{1}{l|}{27.97, 0.00}   & \multicolumn{1}{l|}{27.97, 1.47}  & \multicolumn{1}{l|}{27.92, 1.47}  & \multicolumn{1}{l|}{27.97, 1.48}  & 27.93, 1.48  \\
                                        &                            & \multicolumn{1}{l|}{28.08, 0.00}  & \multicolumn{1}{l|}{28.12, 0.00}   & \multicolumn{1}{l|}{28.02, 1.47}  & \multicolumn{1}{l|}{28.02, 1.72}   & \multicolumn{1}{l|}{28.13, 0.00}  & \multicolumn{1}{l|}{28.14, 0.00}  & \multicolumn{1}{l|}{28.43, 0.29}  & \multicolumn{1}{l}{28.08, 0.00}  \\
                                        &                            & \multicolumn{1}{l|}{28.44, 0.20}  & \multicolumn{1}{l|}{28.44, 0.21}   & \multicolumn{1}{l|}{28.44, 0.31}  & \multicolumn{1}{l|}{28.44, 0.40}   & \multicolumn{1}{l|}{28.44, 0.21}  & \multicolumn{1}{l|}{28.44, 0.20}  & \multicolumn{1}{l|}{30.30, 0.55}  & 28.45, 0.27  \\
                                        &                            & \multicolumn{1}{l|}{30.31, 0.50}  & \multicolumn{1}{l|}{30.30, 0.55}   & \multicolumn{1}{l|}{30.27, 0.75}  & \multicolumn{1}{l|}{30.20, 0.82}   & \multicolumn{1}{l|}{30.30, 0.68}  & \multicolumn{1}{l|}{30.30, 0.57}  & \multicolumn{1}{l|}{30.78, 0.42}  & 30.33, 0.52  \\
                                        &                            & \multicolumn{1}{l|}{30.78, 0.37}  & \multicolumn{1}{l|}{30.77, 0.34}   & \multicolumn{1}{l|}{30.81, 0.51}  & \multicolumn{1}{l|}{30.81, 0.55}   & \multicolumn{1}{l|}{30.78, 0.34}  & \multicolumn{1}{l|}{30.77, 0.35}  & \multicolumn{1}{l|}{32.04, 0.66}  & 30.79, 0.44  \\
                                        &                            & \multicolumn{1}{l|}{32.07, 0.57}  & \multicolumn{1}{l|}{32.11, 0.49}   & \multicolumn{1}{l|}{32.02, 0.80}  & \multicolumn{1}{l|}{31.92, 0.90}   & \multicolumn{1}{l|}{32.13, 0.59}  & \multicolumn{1}{l|}{32.08, 0.46}  & \multicolumn{1}{l|}{34.88, 0.48}  & 32.15, 0.48  \\
                                        &                            & \multicolumn{1}{l|}{34.91, 0.53}  & \multicolumn{1}{l|}{34.89, 0.38}   & \multicolumn{1}{l|}{34.80, 0.75}  & \multicolumn{1}{l|}{34.87, 0.60}   & \multicolumn{1}{l|}{34.89, 0.54}  & \multicolumn{1}{l|}{34.89, 0.38}  & \multicolumn{1}{l|}{39.00, 8.03}  & 34.95, 0.52  \\
                                        &                            & \multicolumn{1}{l|}{38.55, 9.97}  & \multicolumn{1}{l|}{38.18, 8.89}   & \multicolumn{1}{l|}{38.76, 8.15}  & \multicolumn{1}{l|}{38.59, 7.00}   & \multicolumn{1}{l|}{39.17, 8.43}  & \multicolumn{1}{l|}{38.39, 8.85}  & \multicolumn{1}{l|}{39.31, 0.63}  & 39.14, 9.89  \\
                                        &                            & \multicolumn{1}{l|}{39.33, 0.49}  & \multicolumn{1}{l|}{39.30, 0.56}   & \multicolumn{1}{l|}{39.30, 0.78}  & \multicolumn{1}{l|}{39.21, 0.91}   & \multicolumn{1}{l|}{39.36, 0.44}  & \multicolumn{1}{l|}{39.35, 0.53}  & \multicolumn{1}{l|}{46.95, 9.97}  & 39.40, 0.52  \\
                                        &                            & \multicolumn{1}{l|}{47.39, 9.60}  & \multicolumn{1}{l|}{47.27, 8.67}   & \multicolumn{1}{l|}{47.58, 8.66}  & \multicolumn{1}{l|}{47.28, 8.26}   & \multicolumn{1}{l|}{47.15, 7.92}  & \multicolumn{1}{l|}{47.18, 8.45}  & \multicolumn{1}{l|}{47.03, 0.38}  & 47.06, 9.49 \\
                                        &                            & \multicolumn{1}{l|}{52.50, 1.19}  & \multicolumn{1}{l|}{52.46, 1.06}   & \multicolumn{1}{l|}{52.42, 1.35}  & \multicolumn{1}{l|}{52.28, 1.67}   & \multicolumn{1}{l|}{52.50, 1.12}  & \multicolumn{1}{l|}{52.49, 0.99}  & \multicolumn{1}{l|}{52.46, 1.07}  & 52.55, 1.02 \\
                                        &                            & \multicolumn{1}{l|}{57.83, 2.16}  & \multicolumn{1}{l|}{58.04, 3.02}   & \multicolumn{1}{l|}{57.96, 3.89}  & \multicolumn{1}{l|}{57.84, 3.65}   & \multicolumn{1}{l|}{57.59, 4.14}  & \multicolumn{1}{l|}{58.42, 3.45}  & \multicolumn{1}{l|}{57.94, 5.05}  & 57.93, 4.74 \\
                                        &                            & \multicolumn{1}{l|}{65.24, 7.95}  & \multicolumn{1}{l|}{65.28, 7.65}   & \multicolumn{1}{l|}{65.23, 6.51}  & \multicolumn{1}{l|}{65.38, 6.42}   & \multicolumn{1}{l|}{65.18, 7.31}  & \multicolumn{1}{l|}{64.88, 6.04}  & \multicolumn{1}{l|}{65.12, 5.91}  & 65.20, 6.03  \\
                                        &                            & \multicolumn{1}{l|}{66.93, 0.28}  & \multicolumn{1}{l|}{66.92, 0.34}   & \multicolumn{1}{l|}{66.98, 0.68}  & \multicolumn{1}{l|}{66.82, 0.87}   & \multicolumn{1}{l|}{67.00, 0.19}  & \multicolumn{1}{l|}{67.01, 0.56}  & \multicolumn{1}{l|}{67.01, 0.57}  & 67.05, 0.44  \\
                                        &                            & \multicolumn{1}{l|}{74.99, 2.83}  & \multicolumn{1}{l|}{75.42, 3.62}   & \multicolumn{1}{l|}{74.98, 3.68}  & \multicolumn{1}{l|}{75.14, 3.98}   & \multicolumn{1}{l|}{74.78, 2.84}  & \multicolumn{1}{l|}{75.28, 4.44}  & \multicolumn{1}{l|}{75.21, 4.31}  & 75.30, 4.32 \\
                                        &                            & \multicolumn{1}{l|}{76.75, 0.30}  & \multicolumn{1}{l|}{77.72, 3.26}   & \multicolumn{1}{l|}{77.72, 3.26}  & \multicolumn{1}{l|}{79.14, 3.91}   & \multicolumn{1}{l|}{76.72, 0.53}  & \multicolumn{1}{l|}{81.26, 0.27}  & \multicolumn{1}{l|}{82.36, 3.24}  & 82.76, 3.12 \\
                                        &                            & \multicolumn{1}{l|}{85.50, 8.06}  & \multicolumn{1}{l|}{85.87, 11.68}  & \multicolumn{1}{l|}{85.89, 4.82}  & \multicolumn{1}{l|}{86.66, 8.63}   & \multicolumn{1}{l|}{84.62, 7.19}  & \multicolumn{1}{l|}{85.45, 4.12}  & \multicolumn{1}{l|}{85.26, 2.25}  & 85.04, 3.96  \\
                                        &                            & \multicolumn{1}{l|}{87.83, 0.51}  & \multicolumn{1}{l|}{87.81, 0.38}   & \multicolumn{1}{l|}{87.77, 1.08}  & \multicolumn{1}{l|}{87.79, 0.78}   & \multicolumn{1}{l|}{87.83, 0.55}  & \multicolumn{1}{l|}{87.87, 0.52}  & \multicolumn{1}{l|}{87.84, 0.67}  & 87.89, 0.63  \\
                                        &                            & \multicolumn{1}{l|}{92.66, 0.36}  & \multicolumn{1}{l|}{92.71, 0.27}   & \multicolumn{1}{l|}{92.48, 0.95}  & \multicolumn{1}{l|}{92.76, 0.73}  & \multicolumn{1}{l|}{92.66, 0.36}  & \multicolumn{1}{l|}{92.71, 0.38}  & \multicolumn{1}{l|}{92.69, 0.51}  & 92.72, 0.50 \\
                                        &                            & \multicolumn{1}{l|}{105.36, 0.14} & \multicolumn{1}{l|}{105.40, 0.67}  & \multicolumn{1}{l|}{105.19, 0.71} & \multicolumn{1}{l|}{105.44, 1.54} & \multicolumn{1}{l|}{105.35, 0.14} & \multicolumn{1}{l|}{105.40, 0.02} & \multicolumn{1}{l|}{105.35, 0.20} & 105.42, 0.06 \\
                                        &                            & \multicolumn{1}{l|}{122.74, 1.22} & \multicolumn{1}{l|}{127.27, 19.37} & \multicolumn{1}{l|}{122.05, 1.39} & \multicolumn{1}{l|}{125.46, 18.57}  & \multicolumn{1}{l|}{122.33, 1.11} & \multicolumn{1}{l|}{122.41, 0.46} & \multicolumn{1}{l|}{122.93, 2.12} & 122.25, 0.17 \\
                    \multicolumn{1}{l|}{}                   &                            & \multicolumn{1}{l|}{144.57, 0.60} & \multicolumn{1}{l|}{144.66, 0.80}  & \multicolumn{1}{l|}{144.26, 1.43} & \multicolumn{1}{l|}{144.65, 1.55}              & \multicolumn{1}{l|}{144.59, 0.50} & \multicolumn{1}{l|}{144.71, 0.16} & \multicolumn{1}{l|}{144.56, 0.43}       & 144.80, 0.24             \\ \hline
\multirow{3}{*}{Int. cont.} & \multicolumn{1}{c|}{$\Gamma_1$} & \multicolumn{8}{c}{-0.047} \\
 & \multicolumn{1}{c|}{$\Gamma_2$} & \multicolumn{8}{c}{-0.838} \\
 & \multicolumn{1}{c|}{$E_\text{break}$} & \multicolumn{8}{c}{121.86 keV} \\
\hline
\multicolumn{1}{l|}{}                   &  
\multicolumn{1}{c|}{$N_\text{H}$}               & \multicolumn{8}{c}{$4.7\times 10^{20}\text{\,cm}^{-2}$}                                                         \\

\multicolumn{1}{c|}{CXB}                &  \multicolumn{1}{c|}{$\mathcal{F}_\text{3--20\,keV}^\text{CXB}$}     & \multicolumn{8}{c}{$2.82 \times 10^{-11} \text{\,erg\,cm}^{-2}\text{\,s}^{-1}\text{\,deg}^{-2}$}                                                                                                                                                                                                                                           \\
\multicolumn{1}{l|}{}                   &  
\multicolumn{1}{c|}{$E_\text{fold}$} & \multicolumn{8}{c}{41.13 keV}                                                                                                                                                                                                                                            \\

\multicolumn{1}{l|}{}                   &  
\multicolumn{1}{c|}{$\Gamma_\text{CXB}$} & \multicolumn{8}{c}{1.29}                                                                                                                                                                                                                                            \\

\hline

\multirow{6}{*}{Solar}                   &     \multicolumn{1}{c|}{$\Gamma_\text{sol}$}                       & \multicolumn{1}{c|}{8.37}             & \multicolumn{1}{c|}{9.50}              & \multicolumn{1}{c|}{8.90}             & \multicolumn{1}{c|}{8.10}              & \multicolumn{1}{c|}{9.49}             & \multicolumn{1}{c|}{9.50}             & \multicolumn{1}{c|}{9.50}             &    \multicolumn{1}{c}{9.50}          \\
\multicolumn{1}{c|}{}                   &       \multicolumn{1}{c|}{$E_0$, FWHM}                     & \multicolumn{1}{l|}{3.77, 0.00}             & \multicolumn{1}{l|}{3.77, 0.00}              & \multicolumn{1}{l|}{3.77, 0.00}             & \multicolumn{1}{l|}{3.77, 0.00}              & \multicolumn{1}{l|}{3.72, 0.00}             & \multicolumn{1}{l|}{3.72, 0.04}             & \multicolumn{1}{l|}{3.84, 0.00}             &      \multicolumn{1}{l}{3.77, 0.00} \\
\multicolumn{1}{c|}{}                   &    \multicolumn{1}{c|}{$E_0$, FWHM}       
& \multicolumn{1}{l|}{4.45, 0.48}      & \multicolumn{1}{l|}{4.46, 0.56}              & \multicolumn{1}{l|}{4.46, 0.56}             & \multicolumn{1}{l|}{4.46, 0.57}              & \multicolumn{1}{l|}{4.42, 0.32}             & \multicolumn{1}{l|}{4.46, 0.57}             & \multicolumn{1}{l|}{4.45, 0.55}             &      \multicolumn{1}{l}{4.40, 0.62} \\
\multicolumn{1}{c|}{}                   &      \multicolumn{1}{c|}{$E_0$, FWHM}     
& \multicolumn{1}{l|}{5.33, 0.63}      & \multicolumn{1}{l|}{5.33, 0.62}              & \multicolumn{1}{l|}{5.33, 0.62}             & \multicolumn{1}{l|}{5.33, 0.62}              & \multicolumn{1}{l|}{5.32, 0.74}             & \multicolumn{1}{l|}{5.33, 0.63}             & \multicolumn{1}{l|}{5.34, 0.62}             &      \multicolumn{1}{l}{5.35, 0.45} \\
\multicolumn{1}{c|}{}                   &      \multicolumn{1}{c|}{$E_0$, FWHM}     
& \multicolumn{1}{l|}{5.79, 0.00}      & \multicolumn{1}{l|}{5.80, 0.00}              & \multicolumn{1}{l|}{5.80, 0.00}             & \multicolumn{1}{l|}{5.80, 0.00}              & \multicolumn{1}{l|}{5.91, 0.00}             & \multicolumn{1}{l|}{5.80, 0.00}             & \multicolumn{1}{l|}{5.84, 0.00}             &      \multicolumn{1}{l}{5.80, 0.00} \\
\multicolumn{1}{c|}{}                   &   \multicolumn{1}{c|}{$E_0$, FWHM}        
& \multicolumn{1}{l|}{6.48, 0.22}      & \multicolumn{1}{l|}{6.48, 0.23}              & \multicolumn{1}{l|}{6.48, 0.23}             & \multicolumn{1}{l|}{6.48, 0.23}              & \multicolumn{1}{l|}{6.49, 0.13}             & \multicolumn{1}{l|}{6.48, 0.25}             & \multicolumn{1}{l|}{6.46, 0.35}             &      \multicolumn{1}{l}{6.47, 0.29} \\

\end{tabular}
\caption{Parameters for the background model of Sec.~\ref{sec:parametric_model}. ``Int. lines'' and ``Int. cont.'' refer to the internal detector lines and continuum, respectively. Line centroids $E_0$ and FWHM are given in keV. We provide two decimal places for the line $E_0$ and FWHM for numerical reasons, and are not indicative of precision.}
\label{table:parametric_model}

\end{table*}

\pagebreak

\section{\label{sec:appB}{Parametric Background Analysis Fits}}
\renewcommand{\thefigure}{\Alph{section}\arabic{figure}}

\begin{figure}[h!]
    \centering
    \hspace*{-1cm}
    \includegraphics[scale=0.80]{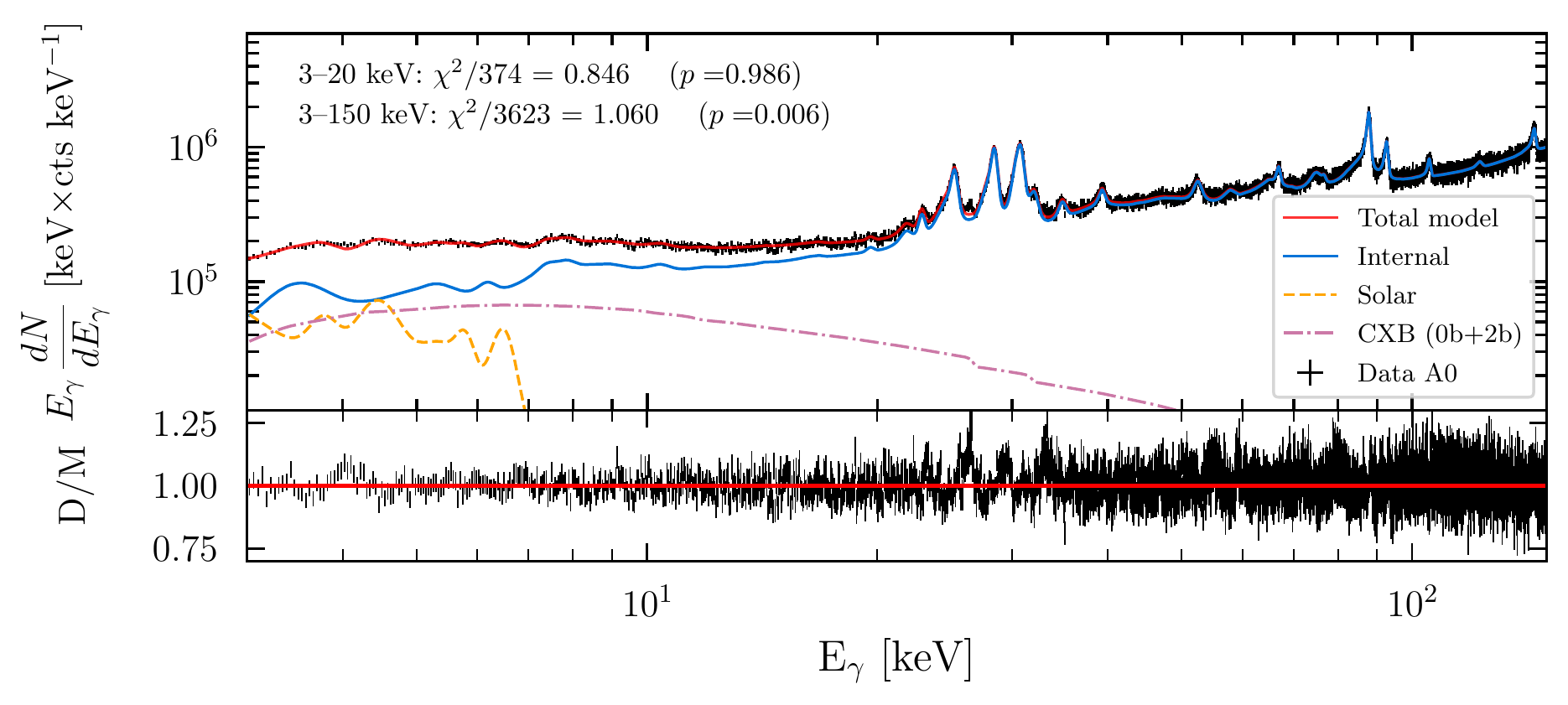}
    \caption{Spectrum and best-fit parametric background model for detector A0 described in Sec.~\ref{sec:parametric}. The bottom panel shows the ratio Data/Model.}
    \label{fig:parametric_A0}

\end{figure}

\begin{figure}[h!]
    \centering
    \hspace*{-1cm}
    \includegraphics[scale=0.80]{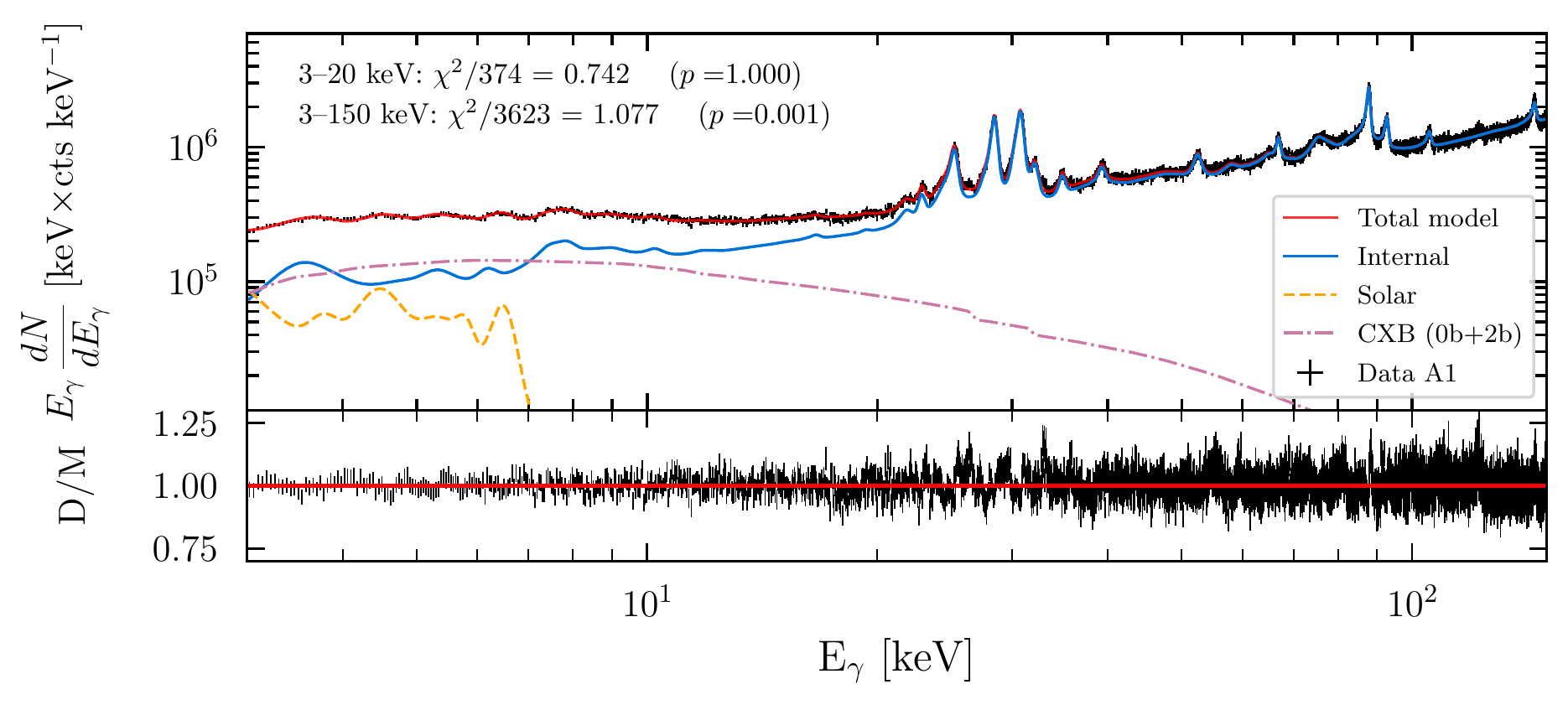}
    \caption{Same as previous, for detector A1.}
    \label{fig:parametric_A1}

\end{figure}

\begin{figure}[h!]
    \centering
    \hspace*{-1cm}
    \includegraphics[scale=0.80]{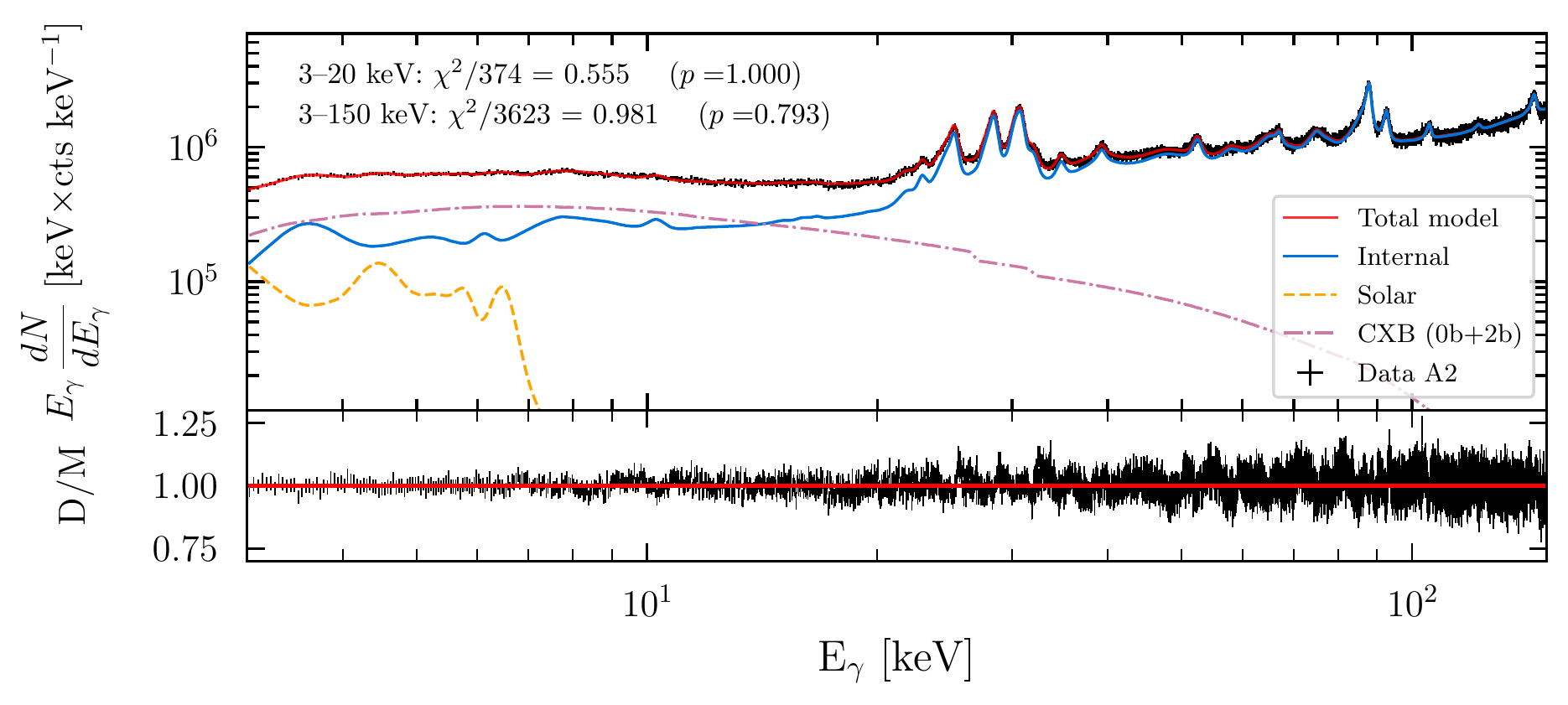}
    \caption{Same as previous, for detector A2.}
    \label{fig:parametric_A2}

\end{figure}

\begin{figure}[h!]
    \centering
    \hspace*{-1cm}
    \includegraphics[scale=0.80]{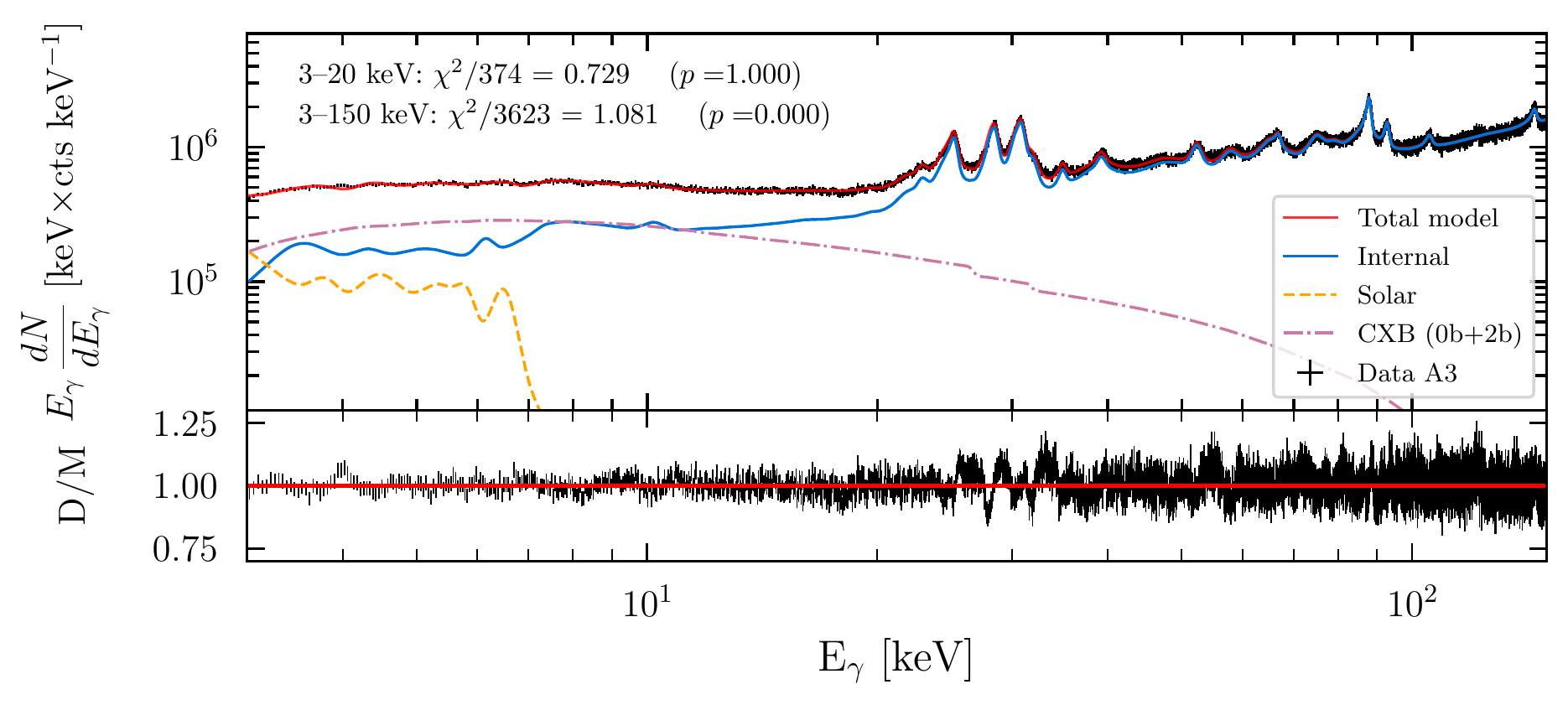}

    \caption{Same as previous, for detector A3.}
    \label{fig:parametric_A3}

\end{figure}

\begin{figure}[h!]
    \centering
    \hspace*{-1cm}
    \includegraphics[scale=0.80]{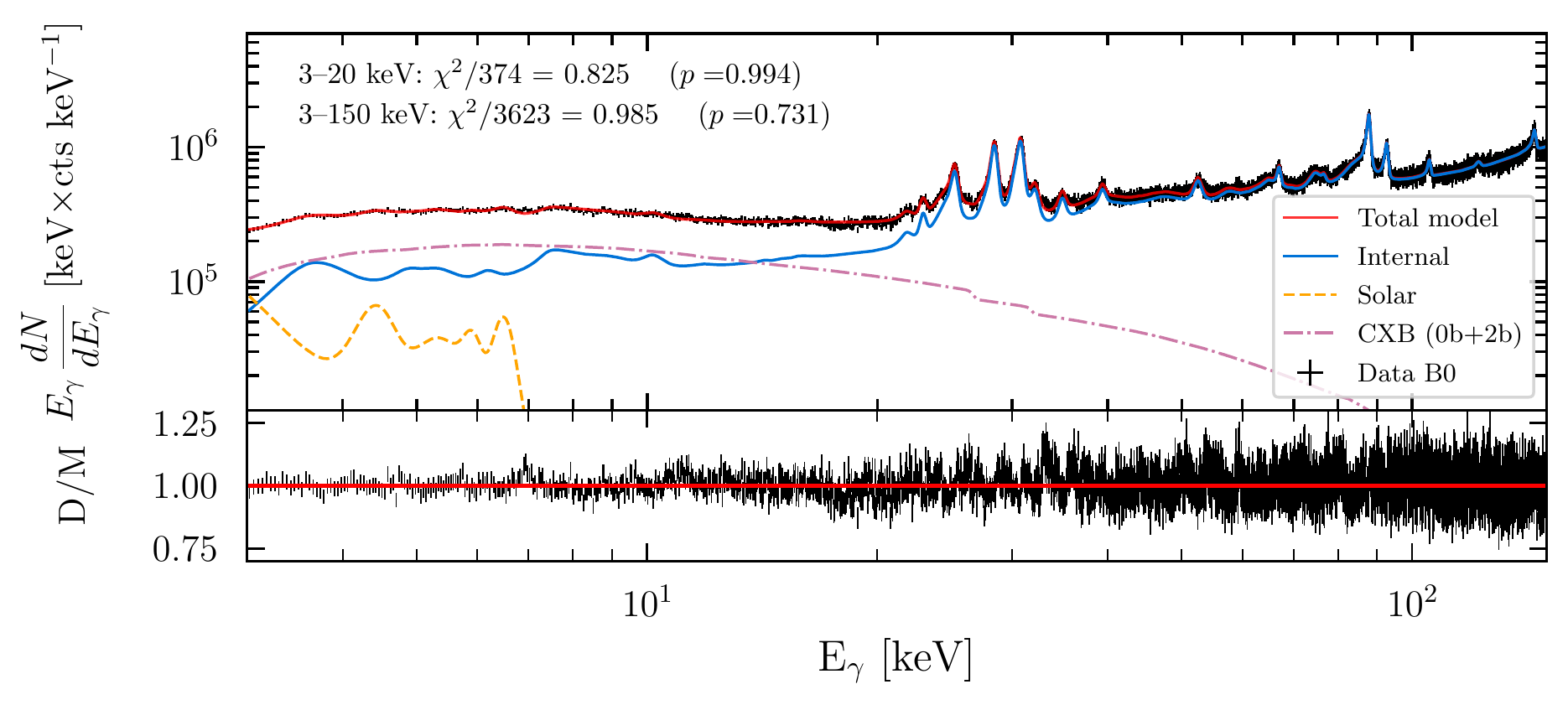}

    \caption{Same as previous, for detector B0.}
    \label{fig:parametric_B0}

\end{figure}

\begin{figure}[h!]
    \centering
    \hspace*{-1cm}
    \includegraphics[scale=0.80]{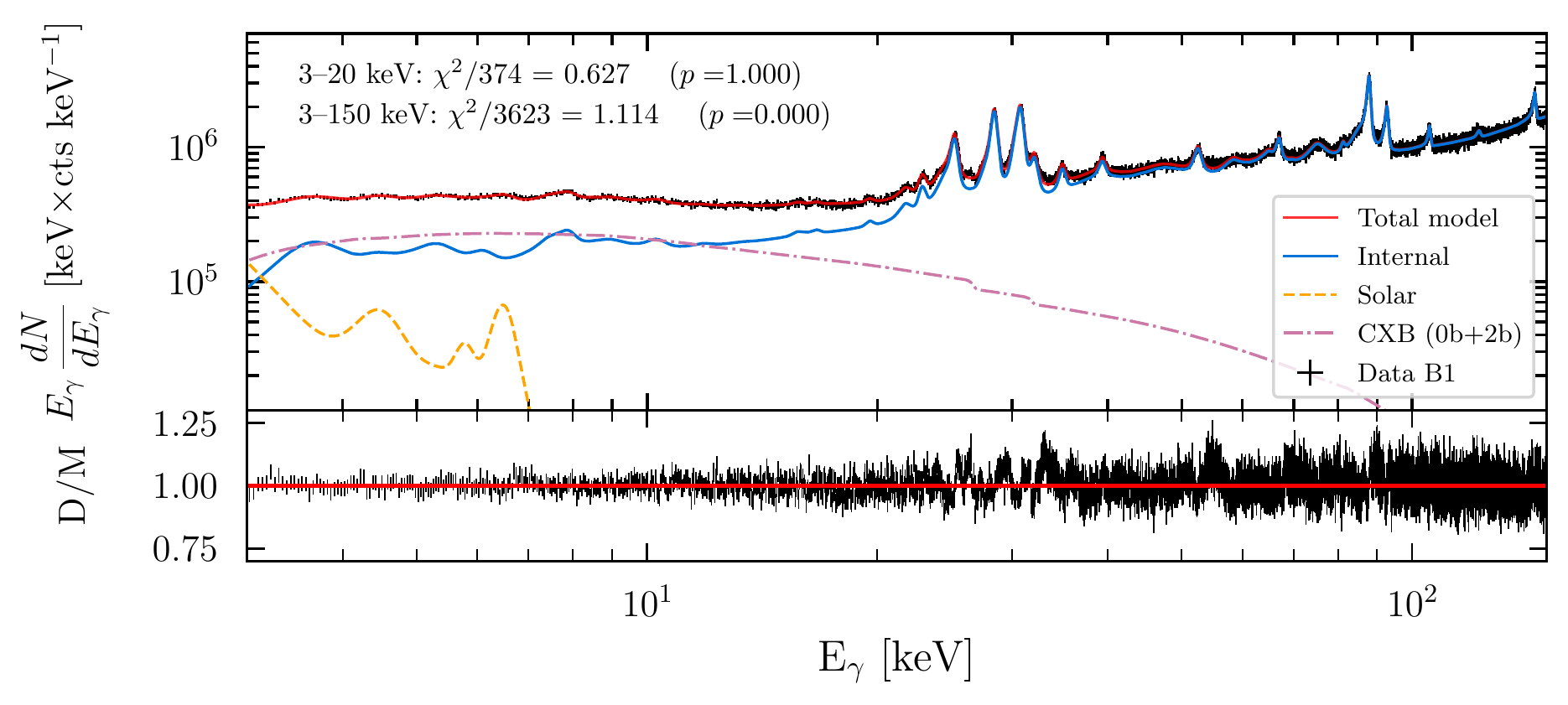}

    \caption{Same as previous, for detector B1.}
    \label{fig:parametric_B1}

\end{figure}

\begin{figure}[h!]
    \centering
    \hspace*{-1cm}
    \includegraphics[scale=0.80]{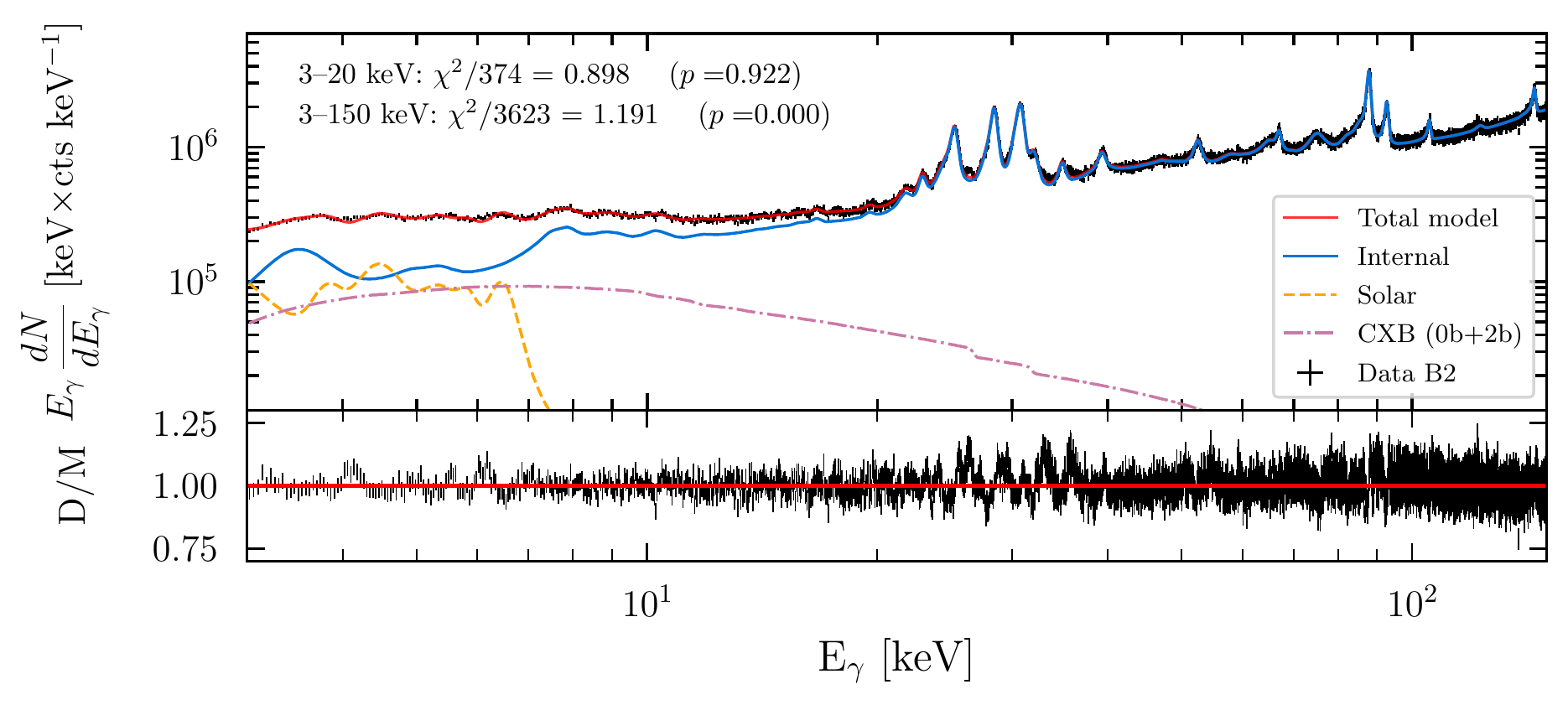}

    \caption{Same as previous, for detector B2.}
    \label{fig:parametric_B2}

\end{figure}

\begin{figure}[h!]
    \centering
    \hspace*{-1cm}
    \includegraphics[scale=0.80]{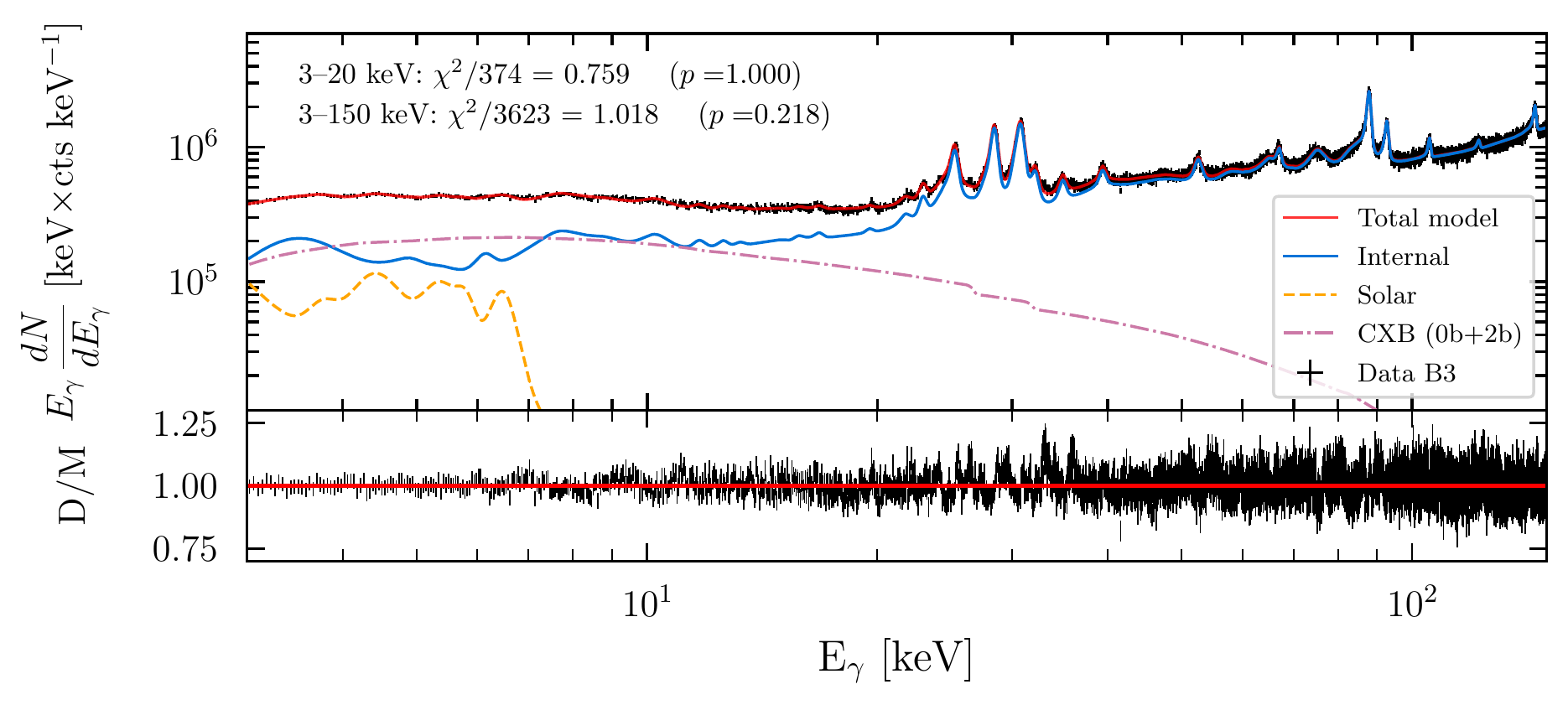}
    \caption{Same as previous, for detector B3.}
    \label{fig:parametric_B3}

\end{figure}

In this section, we show the best fits to the eight individual detector spectra (Figs.~\ref{fig:parametric_A0}-\ref{fig:parametric_B3}) using the parametric background model described in Sec.~\ref{sec:parametric} in the energy range 3--150 keV. As shown, the fit quality in the energy range 3--20 keV is excellent, a result of the improved background modeling as well as the 2.5\% systematic. The fit quality in the full 3--150-keV energy range is somewhat worse, resulting mainly from deviations around the bright detector fluorescence/activation lines between 20--40 keV. In both energy ranges, the $\chi^2$ statistic is calculated with all detector line normalizations free.

\end{document}